\documentclass[twocolumn,,amsmath,amssymb]{revtex4}

\usepackage{gensymb}
\usepackage{graphicx}
\usepackage{dcolumn}
\usepackage{bm}

\newcommand{\be}{\begin{equation}}
\newcommand{\ee}{\end{equation}}
\newcommand{\bfig}{\begin{figure}}
\newcommand{\efig}{\end{figure}}

\begin{document}      
\title{Orthogonal Magnetization and Symmetry Breaking in Pyrochlore Iridate Eu$_2$Ir$_2$O$_7$
} 

\author{Tian Liang$^{1,\ast}$, Timothy H. Hsieh$^{2,\ast,\dagger}$, Jun J. Ishikawa$^3$, Satoru Nakatsuji$^{3,4}$, Liang Fu$^2$ and N. P. Ong$^1$}
\affiliation{
$^1$Department of Physics, Princeton University, Princeton, NJ 08544\\
$^2$Department of Physics, Massachusetts Institute of Technology, Cambridge, MA 02139\\
$^3$Institute for Solid State Physics, University of Tokyo, Kashiwa 277-8581, Japan\\
$^4${PRESTO, Japan Science and Technology Agency (JST), 4-1-8 Honcho Kawaguchi, Saitama 332-0012, Japan}
}

\begin{abstract}
\end{abstract}
 
\maketitle      
{\bf 
Electrons in the pyrochore iridates experience a large interaction energy in addition to a strong spin-orbit interaction. Both features make the iridates promising for realizing novel states such as the Topological Mott Insulator. The pyrochlore iridate Eu$_2$Ir$_2$O$_7$ shows a metal-insulator transition at $T_N \sim$ 120 K below which a magnetically ordered state develops. Using torque magnetometry, we uncover a highly unusual magnetic response. A magnetic field $\bf H$ applied in its $a$-$b$ plane produces a nonlinear magnetization $M_\perp$ orthogonal to the plane. $M_\perp$ displays a $d$-wave field-angle pattern consistent with octupolar order, with a handedness dictated by field cooling, leading to symmetry breaking of the chirality $\omega$. A surprise is that the lobe orientation of the $d$-wave pattern is sensitive to the direction of the field when the sample is field-cooled below $T_N$, suggestive of an additional order parameter $\eta$ already present at 300 K.
}

\begin{figure*}[t]
\includegraphics[width=16 cm]{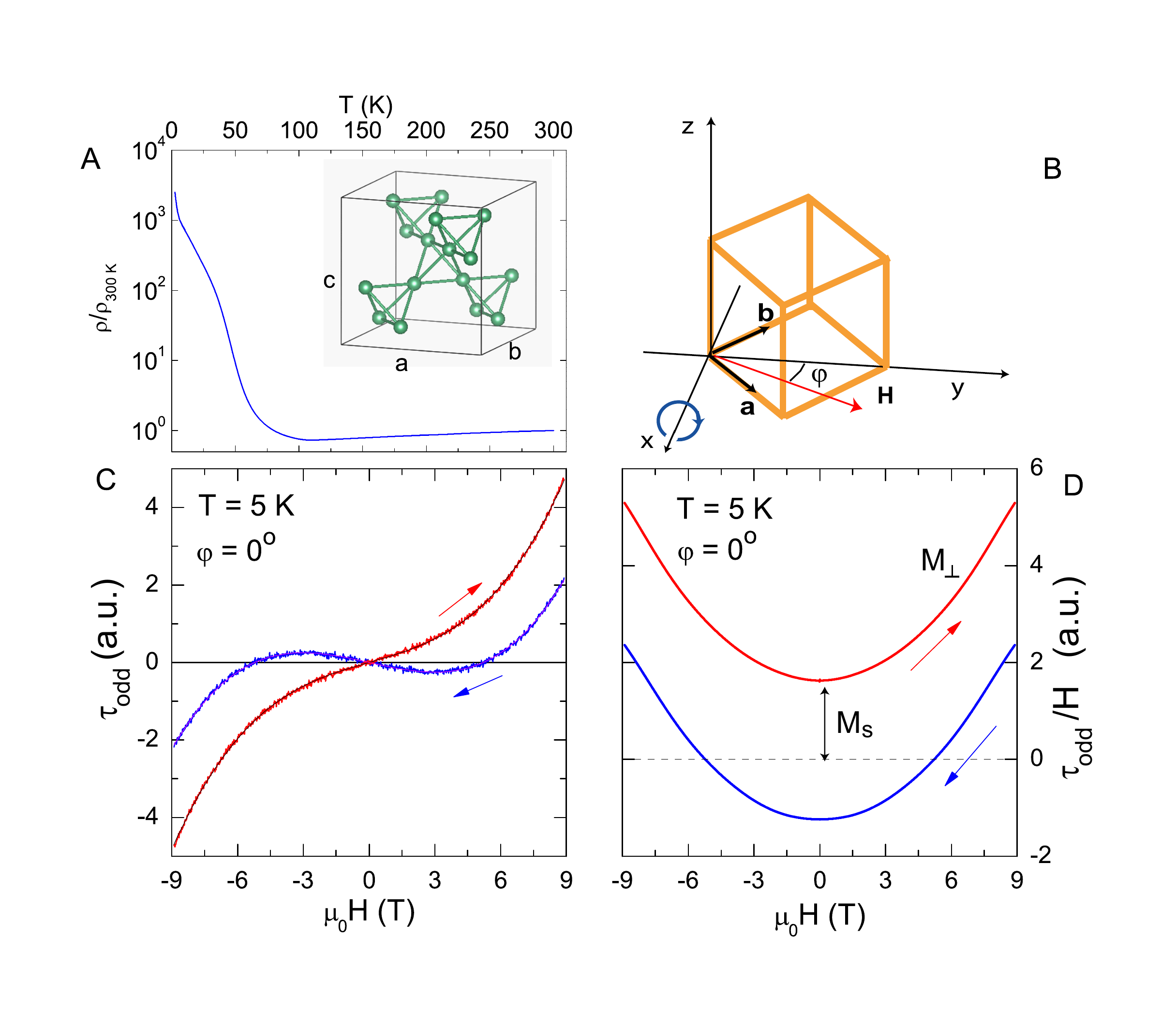}
\caption{\label{figExpt} The pyrochlore lattice, orientation of the lab axes, and analysis of the torque signal.
Panel A shows the lattice of the pyrochlore Eu$_2$Ir$_2$O$_7$. A sketch of the orientation of the lab frame $\bf\hat{x}$, $\bf\hat{y}$ and $\bf\hat{z}$ relative to the lattice vectors $\bf a$, $\bf b$ and $\bf c$ is shown in Panel B. The axis of the torque cantilever is parallel to $\bf\hat{x}$ as indicated by the blue circle. With $\bf H$ (red arrow) in the $a$-$b$ plane at an angle $\varphi$ to $\bf\hat{y}$, the torque component detected is $\tau_{x} = M_{z}H\cos\varphi$.  
Panel C plots the $H$-odd component of the torque $\tau_{odd}$ measured in field sweep-up (-9 $\to$ 9 T, red) and sweep-down (9 $\to$-9 T, blue) scans at 5 K and angle $\varphi$ = 0$^\circ$. In Panel D, we plot $\tau_{odd}/H$ which is the sum of the orthogonal magnetization $M_\perp$ (parabolic curve) displaced vertically by a constant term $M_s$. The sign sgn($M_s$) changes with sweep direction. By contrast, sgn($M_\perp$) which identifies $\omega$ is frozen (the parabolas point ``up'' in both sweeps). See text.
}
\end{figure*}

The pyrochlore iridates, comprised of networks of vertex sharing tetrahedra~\cite{Jackeli2009,Balents2010}, have emerged as candidates for investigating the role of interactions in topological matter~\cite{Balents2010,Balents}. The phase diagram is predicted to have topological states with exotic excitations~\cite{Ashvin2011,Ashvin2013,YBKim_Balents_2014,YBKim2012_2,Balents2014}.
At each Ir$^{4+}$ site, the five 5$d$ electrons occupy the 6 $t_{2g}$ orbitals derived from $d_{xy}$, $d_{yz}$ and $d_{zx}$ states (crystal field splitting lifts the $e_g$ orbitals high above the $t_{2g}$ manifold). The large spin orbital interaction (SOI) splits the $t_{2g}$ manifold into a $J = \frac12$ doublet with energy $\lambda$ and a $J = \frac32$ quadruplet with energy $-\lambda/2$~\cite{BJKim2008, BJKim2009,Balents}. At the critical temperature $T_N\sim$ 120 K, Eu$_2$Ir$_2$O$_7$ undergoes a transition to an insulating state (Fig. \ref{figExpt}A)~\cite{Yamada2007,Matsuhira,Nakatsuji2012,TaftiNakatsuji2012} where a magnetically ordered state emerges~\cite{Nakatsuji_X-ray,Nakatsuji2010,Nakatsuji2011,Arima2012,Yamada2012,Tokura_2012,Tokura2014,Shen2015}. 

In a magnetically ordered state, the free energy $F$ and the magnetization $\bm{M}$ of a system can be expanded, up to third order, as follows, viz.,

\begin{eqnarray}
F = - \bm{M}^d\cdot \bm{H} - \chi^p_{ij} H_i H_j - Q_{ij} H_i H_j - \omega_{ijk} H_i H_j H_k  \label{F}\\
M_i = - \partial F/\partial H_i = M^d_i + \chi^p_{ij} H_j + Q_{ij} H_j + \omega_{ijk} H_j H_k \label{M}
\end{eqnarray}
Here $\bm{M}^d$ (vector) is the conventional dipolar order, $Q_{ij}$ (second rank tensor) is the quadrapolar order, $\omega_{ijk}$ (third rank tensor) is the octupolar order, and $\chi_{ij}$ the conventional paramagnetic susceptibility, with $i,j,k$ referring to components along the unit cell vectors ${\bf a,\,b,\,c}$. $\bm{M}^d$, $Q_{ij}$, $\omega_{ijk}$ represent independent order parameters respectively. We note that, whereas $\bm{M}^d$, $Q_{ij}$, $\omega_{ijk}$ all change their sign under inversion or time-reversal (e.g. inverting the field cooling direction), $\chi_{ij}$ (the paramagnetic susceptibility describing the Zeeman effect) is unaffected by these operations. Specifically, under time reversal, the magnetization $\bm{M}$ of the system becomes,

\be
M_i = - M^d_i + \chi^p_{ij} H_j - Q_{ij} H_j - \omega_{ijk} H_j H_k \label{MT}
\ee
Comparison between Eqs.~\ref{M} and~\ref{MT} shows that $\chi_{ij}$ transforms differently from $Q_{ij}$ (as well as $\bm{M}^d$ and $\omega_{ijk}$). 
We also note that for systems with inversion symmetry like Eu$_2$Ir$_2$O$_7$, the quadrapolar term $Q_{ij}$ vanishes ($F \rightarrow F$, $Q_{ij} \rightarrow - Q_{ij}$, $H \rightarrow - H$). Therefore, the free energy $F$ and magnetization $\bm{M}$ for Eu$_2$Ir$_2$O$_7$ can be written as 

\begin{eqnarray}
F &=& - \bm{M}^d\cdot \bm{H} - \chi^p_{ij} H_i H_j - \omega_{ijk} H_i H_j H_k  \label{EF}\\
M_i &=&  M^d_i + \chi^p_{ij} H_j + \omega_{ijk} H_j H_k \label{EM}
\equiv M^d_i + M_p + M_\perp
\end{eqnarray}
The third term of Eq.~\ref{EM}, $M_\perp = \omega_{ijk} H_j H_k$, which we call orthogonal magnetization, directly detects the octupolar order $\omega_{ijk}$. We note that for a conventional antiferromagnet/ferromagnet, only the first two terms of Eq.~\ref{EM} exist and the third term $\omega_{ijk}$ is absent. Therefore, the detection of the orthogonal magnetization $M_\perp = \omega_{ijk} H_j H_k$ is direct evidence for the ``hidden order'' of the system.

Although the magnetization described by Eq.~\ref{EM} is very interesting, no experiment has been reported. 
In our experiments, a field ${\bf H} = (H_a, H_b,0)$ confined to the $a$-$b$ plane produces a nonlinear magnetization ${\bf M}_{\perp}$ normal to the plane. Depending on the field cooling direction, the observed orthogonal magnetization takes the following forms 
\begin{eqnarray}
M_{\perp} =\omega  {\chi}_{\perp}^{y_{fc}} H_a H_b \quad (\pm \text{{\it y}-axis field cooling})    \label{eq:M1}\\
M_{\perp} =\omega  {\chi}_{\perp}^{x_{fc}} H_x H_y \quad (\pm \text{{\it x}-axis field cooling})     \label{eq:M2}
\end{eqnarray}
where the susceptibility $\chi_{\perp}(T)$ describes its $T$ dependence (an additional phenomenological term $\eta$ is discussed below). In our set-up, we define the (lab) $x$ and $y$ axes as rotated by 45$^\circ$ relative to the lattice vectors $\bf a$ and $\bf b$ of the pyrochlore unit cell [${\bf\hat{x}} \parallel [1\bar{1}0]$, ${\bf\hat{y}} \parallel [110]$ and ${\bf\hat{z}} \parallel [001]$] (Fig. \ref{figExpt}B).
We emphasize that the direction of $\bf M_{\perp}$ cannot be inferred \emph{a priori} from the signs of $H_a$ ($H_x$) and $H_b$ ($H_y$). By necessity, its appearance spontaneously breaks a $Z_2$ symmetry (the system spontaneously chooses $\omega$ to be either +1 or -1). 

We contrast our case with the trivial case of Gd$_2$Ti$_2$O$_7$ (with $T_N\sim 1$ K) in which the applied magnetic field distorts the spin configuration to induce a conventional dipolar magnetization ${\bf M}^d = {\bf M}_{trans}$, previously called ``transverse'' magnetization in~\cite{Sanchez2005,Sanchez2006}.  ${\bf M}_{trans}$ does not involve breaking of a $Z_2$ symmetry, but just comes from conventional Zeeman coupling term $- {\bf M}^d\cdot{\bf H}$. Crucially, the suppression of this Zeeman-induced coupling in Eu$_2$Ir$_2$O$_7$ via large exchange energy $J_{eff} \gtrsim T_N \sim 120$ K allows the octupolar response $M_\perp = \omega_{ijk}H_jH_k$ to emerge. See method section for more discussion.  

We now discuss the experimental data of torque magnetometry.
The axis of the torque cantilever is aligned $\parallel\bf\hat{x}$. With $\bf H$ in the $a$-$b$ plane (at an angle $\varphi $ to ${\bf\hat{y}}$), the torque signal is given by $\tau = M_zH_y$ -- the torque detects the magnetization component $M_z$ normal to the plane in which $\bf H$ lies (Supplementary Figs. S2 and S3).

In Eu$_2$Ir$_2$O$_7$, $M_z$ consists of three terms $M_s \propto H^0$, $M_p \propto H$, $M_\perp \propto H^2$, viz. $M_z$ = $M_s + M_p + M_\perp$. Accordingly, the observed torque (with $\varphi $ and $T$ fixed) can be represented as 
\be
\tau \equiv \tau_s + \tau_p + \tau_\perp = \alpha H + \beta H^2 + \gamma H^3.
\label{eq:tau}
\ee
(We refer to the $H$-even and $H$-odd parts as $\tau_{even} \equiv \tau_p = \beta H^2$ and $\tau_{odd} \equiv \tau_s + \tau_\perp = \alpha H + \gamma H^3$, respectively.)  The first term $\alpha H$ corresponds to a field independent magnetization $M_s \equiv \alpha/\cos\varphi $. The second term $\tau_{even}$ -- the largest term in our field range -- comes from a paramagnetic magnetization $M_p \equiv \beta H/\cos\varphi $. In contrast to $M_s$ and $M_p$ which persist up to 300 K, the third term $\gamma H^3$, which onsets below $T_N$ = 120 K, corresponds to the orthogonal magnetization $M_{\perp}\equiv \gamma H^2/\cos\varphi$ arising from the octupolar order.

By antisymmetrization, we can isolate $\tau_{odd}$, which we plot in Fig. \ref{figExpt}C at 5 K.
The plot clearly shows the $H^3$ variation produced by $M_\perp$ (plus a term from $M_s$). Dividing by $H$, we then isolate $M_\perp$ as a parabola displaced vertically by a constant term $M_s$ (Fig. \ref{figExpt}D). We note that the sign of the constant, sgn($M_s$), reverses between sweep-up and -down curves as expected. However, sgn($M_\perp$) and its absolute value (namely, the curvature of the parabola) stay the same, implying completely different origins between $M_\perp$ and $M_s$ (as well as $M_p$). The contrast sharply excludes the possibility that the orthogonal magnetization $M_\perp$ comes from contamination by the dipolar term $\bm {M}^d$. If $M_\perp$ came from the dipolar term $\bm {M}^d$, it would have shared the same hysteresis patterns of $M_s$. Below, we identify that $M_s$ is related to the phenomenological $\eta$ term which is already present at 300 K. The striking rigidity of sgn($M_\perp$) implies an unusual domain-wall feature of the octupolar order. The procedure is repeated over selected angles $0< \varphi < 360^\circ$ to isolate the $T$ dependence of $M_\perp$ from 5 K to 300 K.

\begin{figure*}[t]
\includegraphics[width=16 cm]{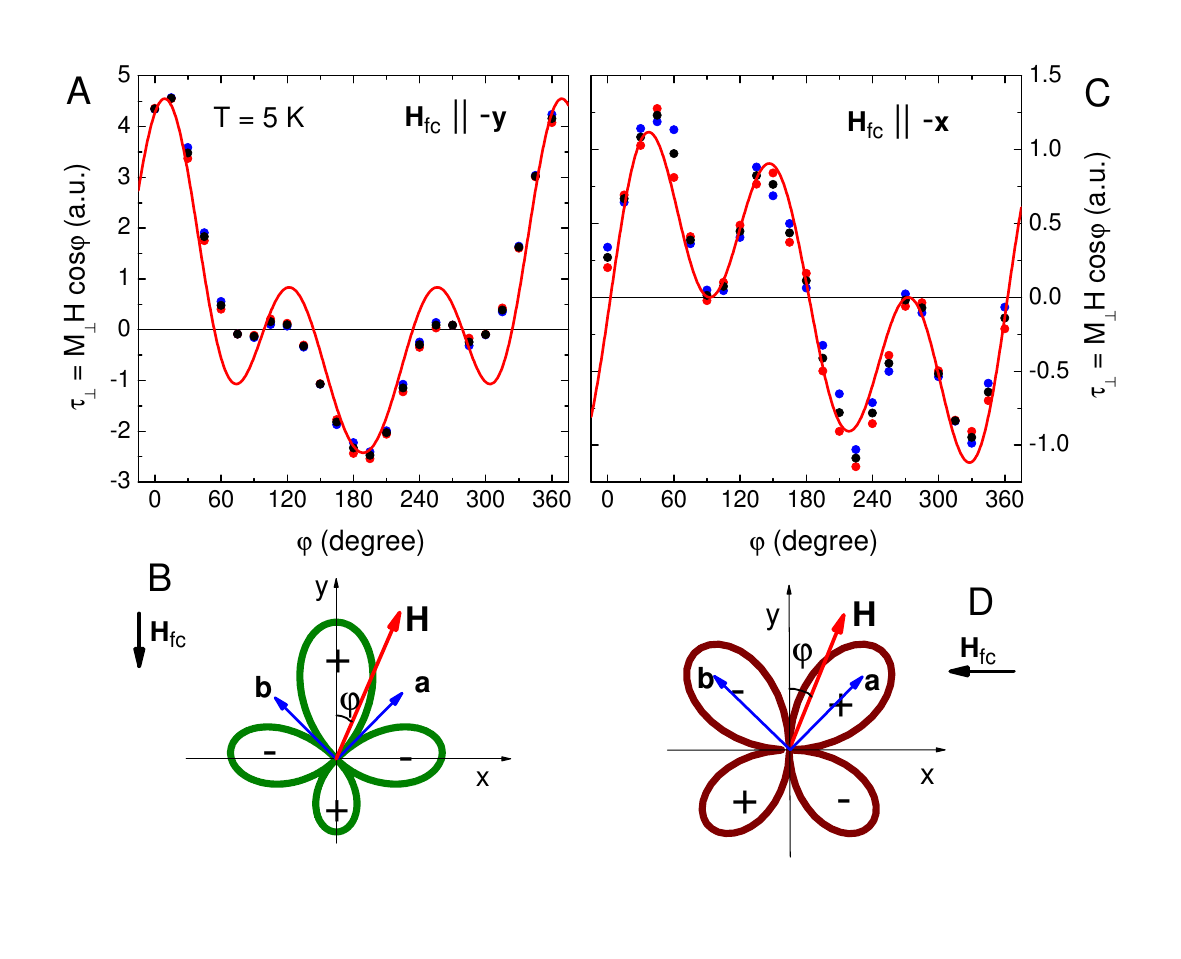}
\caption{\label{figOrtho} The angular variation of $\tau_{\perp} \equiv M_\perp H\cos\varphi$ and the effect of ${\bf H}_{fc}$ on the $d$-wave orientation of $M_\perp$.
Panel A displays $\tau_\perp$ vs. $\varphi$ obtained after cooling in ${\bf H}_{fc}\parallel {-\bf\hat{y}}$. The red curve is the expression $M_\perp\cos\varphi \sim (1+\eta\cos\varphi)\cos 2\varphi\cdot\cos\varphi$, which fits the data well except near 90$^\circ$ and 270$^\circ$. The curve describes the skewed $d$-wave pattern with lobes directed along the $x$ and $y$ axes (Panel B). The signs of the lobes are reversed if the sample is cooled with ${\bf H}_{fc}$ reversed in sign ($\parallel{\bf\hat{y}}$). The experiment is next repeated with ${\bf H}_{fc}$ rotated to $\parallel{-\bf\hat{x}}$. As shown in Panel C, the angular variation of $\tau_\perp$ is now different from Panel A. The data fit well to the expression $(1+\eta\cos\varphi)\sin 2\varphi\cdot\cos\varphi$. The polar representation (Panel D) shows a $d$-wave pattern rotated relative to B, with lobes now directed along the $a$ and $b$ axes. The signs of the lobes are reversed if ${\bf H}_{fc}$ is reversed in sign. In A and C, red (blue) symbols are data obtained in sweep-up (sweep-down) scans while black symbols indicate their average. 
}
\end{figure*}

First, we examine the angular variation of $M_\perp$ at 5 K.
As shown in Fig. \ref{figOrtho}A, the curve of $\tau_\perp = M_\perp H\cos\varphi$ vs. $\varphi$ is plotted. The angular variation is nominally described by the red curve representing a $d$-wave form $\cos 2\varphi$ (Panel B), viz.
\be
M_{\perp}(T,\varphi) = \chi_{\perp}(T)H^2(1+\eta\cos\varphi)\cos 2\varphi,
\label{eq:Mperp}
\ee
where the ``orthogonal'' susceptibility $\chi_{\perp}(T)$ grows like an order parameter below $T_N$. 
The parameter $\eta$, which distorts the $d$-wave pattern, is a phenomenological term that represents an additional order that already exists at 300 K (see below and section \ref{eta} of method section). 
The results in Fig. \ref{figOrtho}A were measured after field-cooling in the (9 T) field ${\bf H}_{fc}\parallel-\bf\hat{y}$. We find that $M_\perp$ changes in sign if ${\bf H}_{fc}$ is inverted. We identify the chirality $\omega = 1$ if ${\bf H}_{fc}\parallel-\bf\hat{y}$ (and -1 if ${\bf H}_{fc}\parallel\bf\hat{y}$). 
This symmetry breaking of the chirality sharply distinguishes the octupolar nature of $M_\perp$ from ``transverse'' magnetization $\bm{M}^d = \bm{M}_{\mathrm{trans}}$ whose origin is strictly dipolar.  

\begin{figure*}[t]
\includegraphics[width=16 cm]{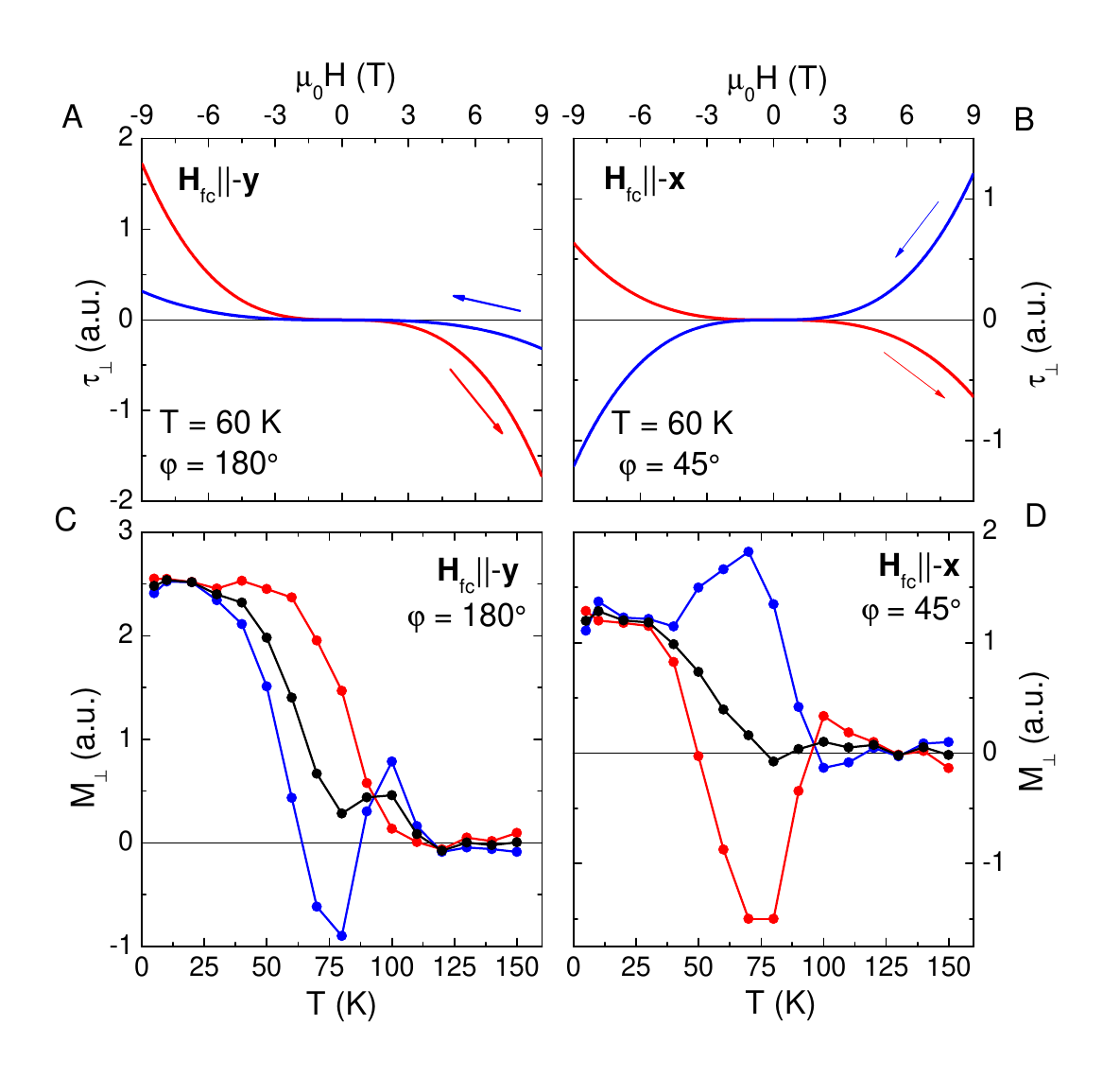}
\caption{\label{figMvsT} The hysteretic behavior and $T$ dependence of the orthogonal magnetization $M_\perp$. 
Panel A displays curves of $\tau_\perp$ measured vs. $H$ at T = 60 K with $\varphi$ fixed at 180$^\circ$ (lobe direction), after cooling in ${\bf H}_{fc} \parallel -{\bf\hat{y}}$. A large hysteresis exists between the field sweep-up (-9 $\to$ 9 T, red curve) and sweep-down (9 $\to$ -9 T, blue) curves. The measurements are repeated with ${\bf H}_{fc}\parallel -{\bf\hat{x}}$. Curves measured in the new lobe direction, $\varphi$ = 45$^\circ$, are shown in Panel B. Panel C shows the $T$ dependence of the orthogonal susceptibility $\chi_\perp\sim M_\perp/H^2$ inferred from sweep-up (red symbols) and sweep-down (blue) curves as shown in Panel A (${\bf H}_{fc}\parallel -{\bf\hat{y}}$). Their average (black symbols) grows like an order parameter below $T_N$. Panel D shows the orthogonal susceptibility vs. $T$ inferred from curves as shown in Panel B. In both C and D, the hysteresis amplitude decreases very rapidly below $\sim$ 80 K, becoming unresolved below 30 K. 
}
\end{figure*}

Further evidence for the octupolar origin of $M_\perp$ derives from the hysteretic behavior of the domain walls (DWs) vs. $T$.
In conventional dipolar magnets, $\bf H$ exerts a strong force on the DW because of dipolar coupling. 
By contrast, for the DW between octupolar domains, a much weaker force is expected. We next describe evidence that the DWs for $M_\perp$ are virtually immobile at low $T$. As already noted in Fig. \ref{figExpt}D, sgn($M_\perp$) is ``frozen'', unlike sgn($M_s$). As $T$ is raised above 25 K, the reversibility gives way to a large hysteresis. At 60 K, $\tau_\perp$ is strongly hysteretic (Figs. \ref{figMvsT}A and \ref{figMvsT}B show the hysteresis observed for the two $d$-wave patterns attained with different ${\bf H}_{fc}$). In Figs. \ref{figMvsT}C and \ref{figMvsT}D, we plot the $T$ dependence of $M_\perp$ measured in up-sweep (red circles) and down-sweep traces (blue) from 5 K to 150 K with $\varphi$ fixed at the lobe maxima.

A striking pattern is that the difference between the red and blue curves (the ``hysteresis amplitude'' $\Delta M_\perp$) is largest near 75 K, but rapidly decreases to below resolution for $T<$ 25 K. This decrease fits well to the thermal activation form $\mathrm{e}^{-\Delta/T}$ with $\Delta$ = 170 and 220 K in C and D, respectively (Supplementary Fig. S6). At each $T$, $\Delta M_\perp$ measures the distance of DW diffusion on our timescales (sweep rates of 1 T/min). Hence the activated form implies diffusion times that grow exponentially with decreasing $T$. The activated form explains why sgn($M_\perp$) is frozen at 5 K in Fig. \ref{figExpt}D. Once a domain pattern is established at 5 K, it is very difficult to erase the pattern because the DWs are immobile on experimental timescales. Both the activated form and the frozen configuration at 5 K reflect the weak coupling of octupolar DWs to $\bf H$. By contrast, the field independent term $M_s$ has a very different hysteretic behavior vs. $T$ (Supplementary Fig. S4).

An unexpected finding is that the angular orientation of the $d$-wave lobes can be rotated by cooling in a field ${\bf H}_{fc}$ parallel to $-\bf\hat{x}$, breaking the underlying lattice symmetry between two ``equivalent'' axes $x$-axis and $y$-axis (the system is cubic). Cooling to 5 K in the new ${\bf H}_{fc}$ leads to the plot of $M_{\perp}$ shown in Fig. \ref{figOrtho}C. The $d$-wave pattern (with $\omega$ = 1) is now shifted by 45$^\circ$ (Fig. \ref{figOrtho}D) and described by 
\be
M_{\perp}(T,\varphi) = \chi_{\perp}(T)H^2(1+\eta\cos\varphi)\sin 2\varphi.
\label{eq:Msin}
\ee 
Here, $\eta$ is again the phenomenological term representing the additional order that already exists at 300 K (see below and section \ref{eta} of method section).
We have also explored cooling with ${\bf H}_{fc}$ in other directions. When cooled in say ${\bf H}_{fc}\parallel\bf a$, the observed $M_{\perp}$ vs. $\varphi$ is a linear combination of the two $d$-wave patterns discussed above. Hence we infer that the two principal axes for field cooling are ${\bf\hat{x}} \parallel (1\bar{1}0)$ and ${\bf\hat{y}} \parallel (110)$ ($\bf\hat{y}$ is identified later as the axis favored by $\eta$). 

The breaking of the underlying $C_4$ lattice symmetry implies that an additional order exists above $T_N$. A first clue comes from the existence of $M_s$ above $T_N$ = 120 K. Figure S10 (SI) shows the angular dependence of $\tau_s$ and $\tau_\perp$ at 150 K. While $\tau_s$ (hence $M_s$) remains finite at 150 K and retains the same angular pattern seen at 5 K (see Fig. S5 in SI), $\tau_\perp$ (hence $M_\perp$) vanishes completely. The differences imply that $M_s$ and $M_\perp$ are associated with very different magnetic orderings. 

\begin{figure*}[t]
\includegraphics[width=16 cm]{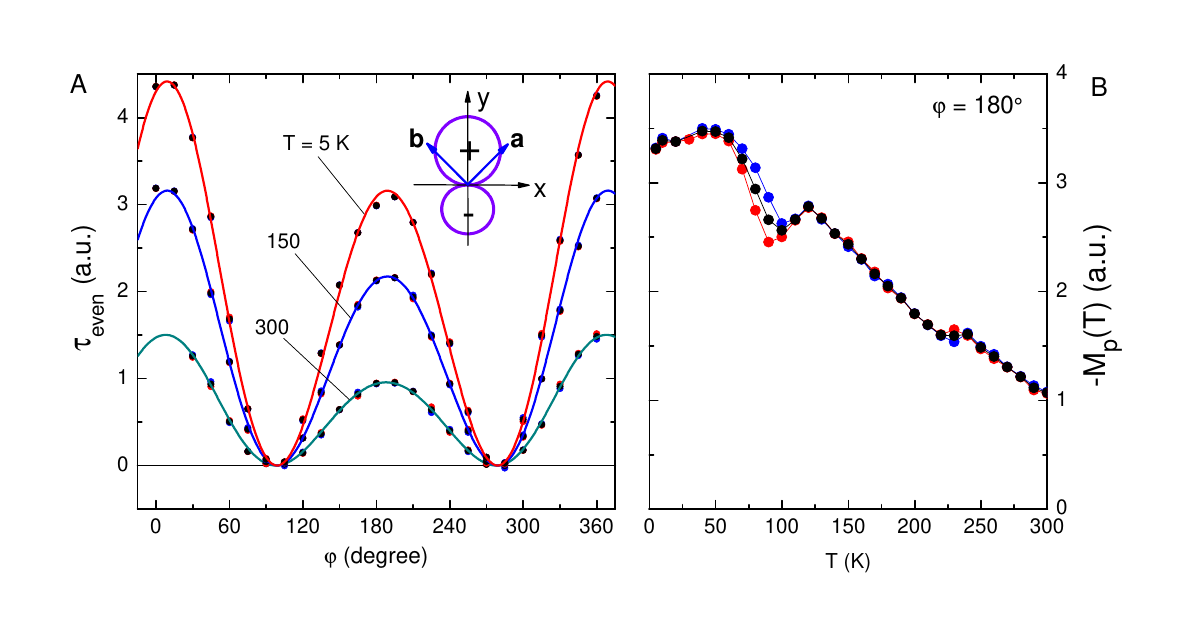}
\caption{\label{figPara} The paramagnetic magnetization. Panel A shows the angular dependence of $H$-even component of the torque $\tau_{even}$ measured at 5, 150 and 300 K. The solid curves are fits to Eq. \ref{eq:Mp}. Amplitude modulation caused by the parameter $\eta$ supresses the maximum at 180$^\circ$ relative to that at 0$^\circ$ (and 360$^\circ$). We estimate that $\eta$ = 0.17 at 5K, 0.18 at 150 K, and 0.22 at 300 K. The suppression is also apparent in the polar representation of the $p$-wave form in $M_p$ shown in the inset. Panel B plots the $T$ dependence of $M_p$ measured at $\varphi  = 180^\circ$. $M_p$ decreases monotonically between 50 K and 300 K, aside from a kink feature near $T_N$ (red and blue circles are $M_p$ measured in sweep up and sweep-down scans; black circles are the average). 
}
\end{figure*}

To investigate this additional order, we examine the paramagnetic term $M_p$ which is strictly $H$-linear, with an angular variation that remains unchanged from 5 to 300 K. Figure \ref{figPara}A plots $\tau_{even}$ ( = $\tau_p$) versus $\varphi$ at $T$ = 5, 150 and 300 K. The sinusoidal variation has the distorted dipolar form (inset) that fits well to the expression $(M_p/H)\cos\varphi$, where $M_p$ has the form
\be
M_p(T,\varphi) = \chi_p(T)H(1+ \eta\cos\varphi)\cos\varphi.
\label{eq:Mp}
\ee
All its $T$ dependence resides in the amplitude $\chi_p(T)$ (Fig. \ref{figPara}B). The parameter $\eta$ (nearly $T$ independent) represents the additional order that develops along $y$-axis, breaking the underlying $C_4$ lattice symmetry.

The symmetry breaking of handedness (chirality $\omega$) together with the activated behavior of DWs of $M_\perp$ sharply distinguish octupolar from dipolar order. The existence of the additional order parameter $\eta$ which already exists at 300 K allows the system to assume two different d-wave lobe patterns of $M_\perp$. Exploring the mechanism of symmetry breaking of handedness in octupolar order, namely, what is the conjugate of the octupolar order parameter, as well as the origin of additional order $\eta$ are fruitful directions to pursue in the iridates.

\cleardoublepage

\section{Methods}

\subsection{Difference between the octupolar and the dipolar order \label{Octupolar}}
In this section, we discuss the difference between the octupolar order and the dipolar order in detail.
As mentioned in the main text, since the Eu$_2$Ir$_2$O$_7$ has inversion symmetry and the quadrapolar order vanishes, the free energy $F$ and the magnetization $\bm{M}$ of the system can in general be expressed as follows, viz.,
\begin{eqnarray}
F &=& - \bm{M}^d\cdot \bm{H} - \chi^p_{ij} H_i H_j - \omega_{ijk} H_i H_j H_k  \label{EF}\\
M_i &=&  M^d_i + \chi^p_{ij} H_j + \omega_{ijk} H_j H_k \label{EM2}\\
&\equiv& M^d_i + M_p + M_\perp \label{O}
\end{eqnarray}
Here $\bm{M}^d$ (vector) is the conventional dipolar order, $\chi_{ij}$ is the conventional paramagnetic susceptibility, and $\omega_{ijk}$ (third rank tensor) is the octupolar order. Accordingly, the magnetization can be written in terms of three terms, the dipolar term $\bm{M}^d$, the paramagnetic term $M_p = \chi^p_{ij} H_j$, and the orthogonal magnetization term $M_\perp = \omega_{ijk} H_j H_k$. While both $\bm{M}^d$ and $\omega_{ijk}$ change sign under time reversal operation, $\chi_{ij}$ stays unchanged. In a conventional antiferromagnet (AF)/ferromagnet (FM), only the first two terms of Eq.~\ref{EM2},~\ref{O} are finite and the octupolar order $\omega_{ijk}$ is absent. Therefore, detection of the orthogonal magnetization $M_\perp = \omega_{ijk}H_jH_k$ is the direct evidence for the ``hidden order'' of the system. 

Our torque magnetometry experiments detect magnetization perpendicular to the applied magnetic field along $z$-axis, $M_z = M_s + M_p + M_\perp$ with $M_s \propto H^0$ the field independent term, $M_p \propto H$ the paramagnetic term, and $M_\perp \propto H^2$ the orthogonal magnetization term.  

Below, we show that the orthognal magnetization $M_\perp$ detected in our experiments comes from the octupolar order $\omega_{ijk}$, and not from the contamination of the conventional dipolar order $\bm{M}^d$.
In the conventional AF/FM, the dipolar magnetization $\bm{M}^d$ is simply represented as the sum of local dipoles $\bm{m}_i$ consisting the system, i.e., $\bm{M}^d = \sum_1^N \bm{m}_i$, with $N$ the total number of the lattice site. 
If $\bm{M}^d = \sum_1^N \bm{m}_i = 0$, then no magnetization can be detected and orthogonal magnetization $M_\perp =0$ rigorously holds. More in general, if $\bm{M}^d = \sum_1^N \bm{m}_i \neq 0$, then in principle $\bm{M}^d$ can appear, if any, in the $M_s$ term of our experiment, and it can even take the highly nonlinear behavior like the case of Gd$_2$Ti$_2$O$_7$ where the trivial ``transverse'' magnetization $\bm{M}^d = \bm{M}_{\mathrm{trans}}$ can appear as a consequece of the distortion of the spin configuration~\cite{Sanchez2005,Sanchez2006}.  However, if this were the case and $M_\perp$ merely came from the contamination of dipolar order $\bm{M}^d$ for Eu$_2$Ir$_2$O$_7$, then the hysteretic behavior of $M_s$ and $M_\perp$ would have been the same because they would have shared the same source $\bm{M}^d$. However, as evidenced in Fig.~1D in the main text, at 5 K, while $M_\perp$ is completely frozen, showing no hysteresis at all, $M_s$ changes sign and manifests a large hysteresis. The angular and temperature dependences of the hysteresis curves of $M_s$ and $M_\perp$ also manifest completely different behaviors as shown in Fig.~S4, ~S5 in the supplement. Furthermore, as mentioned in the main text, while $M_s$ term persists above $T_N \sim 120$ K, and is related to the phenomenological $\eta$ term which is already present at 300 K, orthogonal magnetization $M_\perp$ emerges only below $T_N = 120$ K. These evidences sharply distinguish the different origins between $M_s$ and $M_\perp$, excluding the possibility of contamination of dipolar order $\bm{M}^d$ into orthogonal magnetization $M_\perp$. We also note that while $M_\perp$ and $M_s$ change sign under flippling the field cooling direction, $M_p$ does not, so the origin of $M_\perp$ can easily be separated out from the paramagnetic term $M_p$ as well.

Another way to see that the orthogonal magnetization $M_\perp$ cannot be explained by the contamination of conventional dipolar order $\bm{M}^d$ comes from the comparison of the relevant Zeeman energy scale. We contrast the case of Eu$_2$Ir$_2$O$_7$ where orthogonal magnetization $M_\perp$ appears, with the case of conventional AF Gd$_2$Ti$_2$O$_7$ where the trivial dipolar magnetization, previously called ``transverse'' magnetization $\bm{M}^d = \bm{M}_{\mathrm{trans}}$, appears due to the distortion of spin configuration via the Zeeman energy. 
In Gd$_2$Ti$_2$O$_7$, the relevant macroscopic exchange energy is $J_{\mathrm{eff}} \sim T_N \sim 1$ K (= 0.0866 meV) (microscopic exchange energy is much higher than this) and the magnetic moment is $\sim 7~\mu_B$ (= 7.28 meV at 9 T)~\cite{GdTiOMoment}. Therefore, under applied magnetic field, it is easy to distort the spin configuration to induce the dipolar ``transverse'' magnetization $\bm{M}^d = \bm{M}_{\mathrm{trans}}$ normal to the applied magnetic field. Indeed, in Ref.~\cite{Sanchez2006}, Gd$_2$Ti$_2$O$_7$ shows a sharp kink $\sim 3$ T in torque data, signaling the distortion of the spin configuration.  Above $\sim 3$ T, the local dipoles $\bm{m}_i$ which consist the system, tilt towards the direction of applied magnetic field through the conventional dipolar coupling $-\bm{M}^d\cdot\bm{H}$ to give a highly nonlinear dipolar ``transverse'' magnetization $\bm{M}^d = \bm{M}_{\mathrm{trans}} = \sum_1^N \bm{m}_i$ that cannot be decomposed into the simple polynomial form as seen in Eu$_2$Ir$_2$O$_7$ (see the data for Gd$_2$Ti$_2$O$_7$ in Ref.~\cite{Sanchez2006} for comparison). At sufficiently high enough applied magnetic field, every local magnetic dipoles completely align towards the applied magnetic field, and the  ``transverse'' magnetization $\bm{M}^d = \bm{M}_{\mathrm{trans}}$ vanishes completely. By contrast, in the case of Eu$_2$Ir$_2$O$_7$, the relevant macroscopic exchange energy is $J_{\mathrm{eff}} \sim T_N \sim 120$ K (= 10.4 meV) (microscopic exchange energy is much higher than this) and the magnetic moment of iridium ion Ir$^{4+}$ is $<1.1~\mu_B$ (= 1.14 meV at 9 T)~\cite{EuIrOMuonMoment,YIOMoment1,YIOMoment2}. The Zeeman energy induced by the magnetic field is too small to distort the spin configuration under experimentally accessible field up to 9 T. Indeed, our torque data fits to the simple polynomial form $\tau = \alpha H + \beta H^2 + \gamma H^3$ very smoothly, showing that there is no distortion of the spin configuration. This sharply distinguishes the case of Eu$_2$Ir$_2$O$_7$ in which the orthogonal magnetization $M_\perp = \omega_{ijk} H_j H_k$, i.e., octupolar order $\omega_{ijk}$, is observed, from the case of Gd$_2$Ti$_2$O$_7$ in which the trivial dipolar ``transverse'' magnetization $\bm{M}^d = \bm{M}_{\mathrm{trans}}$ is observed.  See section~\ref{Comparison} for more details.

\subsection{Spontaneous Symmetry Breaking of Handedness (Chirality $\omega$) in $\bm{M}_\perp$ \label{Chirality}}

In this section, we discuss the spontaneous symmetry breaking of handedness (chirality $\omega$) in the orthogonal magnetization $M_\perp$ in detail. As shown in previous section~\ref{Octupolar}, the orthogonal magnetization $M_\perp = \omega_{ijk} H_j H_k$ is the thermodynamical manifestation of the octupolar order $\omega_{ijk}$. As shown in the main text, the measured orthogonal magnetization $M_\perp$ can be represented as follows, depending on the field cooling direction, viz.,
\begin{eqnarray}
M_{\perp} =\omega  {\chi}_{\perp}^{y_{fc}} H_a H_b \quad (\pm \text{{\it y}-axis field cooling})    \label{eq:M1}\\
M_{\perp} =\omega  {\chi}_{\perp}^{x_{fc}} H_x H_y \quad (\pm \text{{\it x}-axis field cooling})     \label{eq:M2}
\end{eqnarray} 
The magnetization response is along the direction orthogonal to the plane defined by $\hat{\bm{a}}$ and $\hat{\bm{b}}$ (or equivalently, $\hat{\bm{x}}$ and $\hat{\bm{y}}$), i.e. $\bm{M}$ is in the $\hat{\bm{z}}$ direction. Accordingly, we call it $M_\perp$. 
It should now be apparent that spontaneous symmetry breaking happens. The free energy $F$ in Eq.~\ref{EF} does not dictate whether $M_\perp$ is along $+\hat{\bm{z}}$ or along $-\hat{\bm{z}}$ (both are allowed). However, in response to $\bm{H}$ applied in the $a-b$ plane, the system \textit{spontaneously} selects one direction. If the direction $+\hat{\bm{z}}$ is selected, the chirality $\omega = +1$. One cannot predict \textit{a priori} whether $\bm{H}$ applied strictly in the $a-b$ plane gives rise to an $M_\perp\cdot\hat{\bm{z}}> 0$ or $M_\perp\cdot\hat{\bm{z}}< 0$. The existence of this spontaneous orthogonal magnetization is the central message of our work.

We now contrast the foregoing with a conventional AF where the ``transverse'' magnetization induced by $\bm{H}$ seems to have engendered considerable confusion. The applied $\bm{H}$ couples to individual subunit moments (e.g. on Mn in MnF$_2$) by the Zeeman energy $E_z$. Because the moments cant towards the direction of $\bm{H}$, there is no spontaneous symmetry breaking of the type discussed above. Following convention, we call the magnetization of the two sublattices $\bm{M}_A$ and $\bm{M}_B$ (they are nominally antiparallel). 
First, if $\bm{H} \perp (\bm{M}_A-\bm{M}_B)$, we obtain a canting of both sublattice magnetizations towards $\bm{H}$, leading to a net magnetization ``transverse'' to $(\bm{M}_A-\bm{M}_B)$, namely $\bm{M}_{trans} \perp (\bm{M}_A-\bm{M}_B)$. This is a trivial Zeeman-driven ``transverse'' magnetization whose direction is dictated by $\bm{H}$. On the other hand, if $\bm{H} \parallel (\bm{M}_A-\bm{M}_B)$, the Zeeman response is initially weak. Increasing $\bm{H}$ leads to a spin-flop transition at which $\bm{M}_A-\bm{M}_B$ suddenly aligns perpendicular to $\bm{H}$. Above the spin flop, we again obtain the same ``transverse'' magnetization $\bm{M}_{trans} \perp (\bm{M}_A-\bm{M}_B)$. In both orientations, there is no spontaneous symmetry breaking; $\bm{M}_{trans}$ trivially aligns with $\bm{H}$. As discussed at length in sections \ref{Octupolar}, \ref{Comparison}, in the conventional AF Gd$_2$Ti$_2$O$_7$, such ``transverse'' magnetization has been previously observed (the spin configuration distorts under applied magnetic field $\bm{H}$ and generates the ``transverse'' magnetization, see sections \ref{Octupolar}, \ref{Comparison} for details). Crucially, the suppression of this Zeeman-induced coupling in Eu$_2$Ir$_2$O$_7$ allows the octupolar response to be observed.

Finally, the observed orthogonal magnetizations $M_\perp$ in Eq.~\ref{eq:M1},~\ref{eq:M2} take two d-wave patterns when $\bm{H}$ is rotated in the $a-b$ plane as shown in Fig.~2 of the main text. Such d-wave variation vs. $\varphi$ with alternating signs cannot be produced by the ``transverse'' magnetization in a conventional AF. 

\subsection{Difference between ${\bf M}_\perp$ and ${\bf M}_{trans}$ \label{Comparison}}
In this section, we discuss in some detail how the orthogonal magnetization ${\bf M}_\perp$ observed in this work is distinct from the ``transverse'' magnetization ${\bf M}_{trans}$ previously studied in the pyrochlore magnet Gd$_2$Ti$_2$O$_7$~\cite{Sanchez2005, Sanchez2006}. Their origins and physical implications are very different. (In Refs.~\cite{Sanchez2005, Sanchez2006}, the notation ${\bf M}_\perp$ was used for ``transverse'' magnetization ${\bf M}_{trans}$. Here we reserve ${\bf M}_\perp$ for our orthogonal magnetization and use ${\bf M}_{trans}$ to represent the ``transverse'' magnetization for clarity.) 


\subsubsection{Review of $M_{trans}$ in references~\cite{Sanchez2005, Sanchez2006} } 
Following Ref.~\cite{Sanchez2005}, we write the Hamiltonian for a pyrochlore magnet Gd$_2$Ti$_2$O$_7$:
\be
\hat{\cal H} = J\sum_{<ij>}\mathbf{S}_i \cdot \mathbf{S}_j +D\sum_i(\mathbf{n}_i\cdot \mathbf{S}_i)^2-\mathbf{H}\cdot \sum_i\mathbf{S}_i .\label{Hamiltonian}
\ee

The first term in Eq.~\ref{Hamiltonian} is the Heisenberg interaction term with $J>0$. The sum runs over the nearest neighbor sites. The second term is the single-ion interaction term and the third term is the Zeeman coupling term in applied field $\bf H$. $\mathbf{S}_i$ represents the spin on site $i$. $\mathbf{n}_i $ (i = 1-4) are the local easy axes. 

When $D>0$, the single-ion term favors alignment of the spin in the local easy plane normal to ${\bf n}_i$. However, the Heisenberg and Zeeman terms favor $\mathbf{S}_{tet}\parallel\mathbf{H}$, where $\mathbf{S}_{tet} = \sum_{i}^{4}\mathbf{S}_i$ is the sum of the 4 spins in each tetrahedron. 
The two conditions can be simultaneously satisfied when the applied $H$ is small, in which case $\mathbf{S}_{tet}$ = $\mathbf{H}/2J$, so no ``transverse'' magnetization $M_{trans}$ appears. However, since the first constraint restricts the maximum possible value of $\mathbf{S}_{tet}^{max}$ to be smaller than the saturation value of $\mathbf{S}_{tet}^{sat} = 4S$, the Zeeman term causes the spins to cant out of the local easy plane when $\bf H$ exceeds $\mathbf{H}_c = 2J\mathbf{S}_{tet}^{max}$. Hence $\mathbf{S}_{tet} \neq \mathbf{H}/2J$, resulting in the appearance of a ``transverse'' magnetization ${\bf M}_{trans}$. With further increase in $H$, each spin fully aligns with $\bf H$ (when $H>H_{sat}$), and ${\bf M}_{trans}$ vanishes.

We note that the direction of ${\bf M}_{trans}$, induced by canting of the spins out of the local easy plane via Zeeman coupling to $\bf H$, is completely dictated by $\bf H$. It is not related to octupolar order, and does not involve spontaneous breaking of a discrete symmetry.

\subsubsection{Orthogonal Magnetization ${\bf M}_\perp$}
Next we describe the orthogonal magnetization ${\bf M}_\perp$ observed in our experiments.

\begin{enumerate}
	
	\item 
	The orthogonal magnetization ${\bf M}_\perp$ which develops below $T_N$ = 120 K is given by Eqs.~\ref{eq:M1},~\ref{eq:M2}. We note that it involves the chirality $\omega$ multiplied by a \emph{susceptibility} $\chi$, i.e., $\omega\chi_\perp$. By contrast, the ``transverse'' magnetization ${\bf M}_{trans}$ is the normal component of the magnetization induced by Zeeman term. 
	
	
	A key point of the orthogonal magnetization $M_\perp$ is related to the spontaneous symmetry breaking dictated by the sign of the chirality $\omega$, namely, despite the field cooling along $+y$-axis ($+x$-axis) and $-y$-axis ($-x$-axis) nominally gives no difference, the sign of the order parameter, i.e. chirality $\omega$ changes sign, breaking the symmetry. We emphasize that the configurations of dipoles shown in Fig.~S1 in the supplement only serve as the symmetry constraint of the octupolar order the system can take, and the dipoles themselves are not our focus. In other words, Eqs.~\ref{eq:M1}~\ref{eq:M2} do not come from the canting of the spins; ${\bf M}_\perp$ is not induced by a Zeeman term.     
	
	\item 
	A sharp distinction between ${\bf M}_\perp$ and ${\bf M}_{trans}$ is shown by the hysteretic behavior. As shown in the lower panels of Fig.~3 (main text) and in Fig.~S6 (supplement), ${\bf M}_\perp$ does not show any hysteresis below 30 K, whereas large hysteresis is observed between 30 K and 120 K. This is very different from hysteresis caused by motion of conventional Bloch domain walls. 
	
	\item 
	Separation of ${\bf M}_\perp$ from other terms M$_s$ and M$_p$.
	The observed total magnetization ${\bf M}_{obs}$ is the sum of three terms, viz. ${\bf M}_{obs} = {\bf M}_s + {\bf M}_p + {\bf M}_\perp$. All terms are perpendicular to the applied magnetic field. The only important contribution that emerges from the octupolar magnetic order is ${\bf M}_\perp$. We carefully separated out each contribution. It is worth remarking that ${\bf M}_\perp$ does not arise from Taylor expansion of the field independent magnetization ${\bf M}_s$ or the paramagnetic term ${\bf M}_p$. First, one can separate ${\bf M}_\perp$ from ${\bf M}_p$. While ${\bf M}_\perp$ changes sign if the direction of the field-cooling field ${\bf H}_{fc}$ is inverted, ${\bf M}_p$ does not. This shows that the two terms are distinct. Further, ${\bf M}_s$ is easily distinguished from ${\bf M}_\perp$ term by their qualitatively different hysteretic behavior versus field, angle and temperature, as discussed in Sec. S4 in the supplement. 
	
\end{enumerate}

\subsection{Additional order $\eta$ \label{eta}}

In this section we discuss additional order $\eta$ and its relation to $M_s$, $M_p$, $M_\perp$.

\begin{enumerate}
	
	\item 
	In addition to ${\bf M}_\perp$ related to octupolar order, which develops below $T_N$ = 120 K, another order represented by $\eta$ (nearly temperature independent), develops at least up to 300 K ($\eta$ = 0.22), suggesting the origin of $\eta$ is related to higher energy scale. 
	
	\item 
	The $\eta$ develops along $y$-axis, breaking the underlying lattice symmetry between $y$-axis ([110]-axis) and $x$-axis ([1-10]-axis), the two ``equivalent'' axes if only lattice symmetry is considered. 
	
	\item
	The $\eta$ couples to each of the terms $M_s$, $M_p$, and ${\bf M}_\perp$, both below and above $T_N$.
	
	\begin{enumerate}
		
		\item 
		M$_s$ term (as well as M$_p$ term) remains finite above $T_N$ = 120 K (see Fig.~S10 in the supplement), above which ${\bf M}_\perp$ (namely, the octupolar order) vanishes. This again implies that M$_s$ (and M$_p$) are unrelated to ${\bf M}_\perp$. Since the only order parameter which exists above $T_N$ = 120 K is $\eta$, we speculate M$_s$ is intimately related to $\eta$. 
		
		\item 
		The paramagnetic term M$_p$ is perpendicular to the applied magnetic field and show Curie-Weiss like temperature dependence. In general, the perpendicular paramagnetic term can arise in any anisotropic system, and it itself is a trivial effect. 
		The paramagnetic term M$_p$ inherits the anisotropy of $\eta$, breaking the underlying lattice symmetry of $x$-axis and $y$-axis. 
		
		\item
		Since $\eta$ persists above $T_N$ = 120 K, above which ${\bf M}_\perp$ vanishes, the origins of $\eta$ is different from ${\bf M}_\perp$. However, the fact that $\eta$ breaks underlying lattice symmetry between $x$-axis and $y$-axis allows ${\bf M}_\perp$ to assume different d-wave patterns depending on whether the field-cooling direction H$_{fc}$ is along $x$-axis (Eq.~\ref{eq:M2}) or $y$-axis (Eq.~\ref{eq:M1}).
		
	\end{enumerate}
	
	\item
	An unusual anisotropy coming from $\eta$ makes the absolute value at $\varphi$ = 0 $\degree$ ($+y$-axis) different from $\varphi$ = 180 $\degree$ ($-y$-axis) for ${\bf M}_\perp$ and M$_p$. This is anomalous, as $+y$-axis and $-y$-axis should be the same except for flipping the definition of the sign of the applied magnetic field. We take this effect phenomenologically by adding the term $\eta \cos \varphi$. 
	
\end{enumerate}


\newpage

\vspace{1cm}\noindent

\vspace{1mm}
$^\ast${These authors contributed equally to this work.}

\vspace{3mm}
$^\dagger${Current address of T.T.H.: Kavli Institute for Theoretical Physics, University of California, Santa Barbara, CA 93106}

\vspace{5mm}\noindent
{\bf Supplementary Information} is available in the online version of the paper.

\vspace{5mm}\noindent
{\bf Acknowledgements} T.L. acknowledges a scholarship from Japan Student Services Organization. N.P.O. acknowledges the support of the U.S. National Science Foundation (Grant DMR 1420541) and the Gordon and Betty Moore Foundation’s EPiQS Initiative through Grant GBMF4539. L.F. and T.H. were supported by DOE Office of Basic Energy Sciences, DE-SC0010526. T.H. thanks the KITP graduate fellowship program. The research at Univ. Tokyo is supported by grants-in-aid (nos. 25707030) and the Program for Advancing Strategic International Networks to Accelerate the Circulation of Talented Researchers (no. R2604) from JSPS, by PRESTO to JST, and grant-in-aid for scientific research on Innovative Areas (Grants No. 15H05882 and No. 15H05883) from MEXT. 

\vspace{5mm}\noindent
{\bf Author Contributions} T.L., T.H.H., L.F. and N.P.O. conceived the idea behind the experiment. T.L. designed the experiment and carried out all the measurements. T.L. and N.P.O. analysed the results with important insights from T.H.H. and L.F. The manuscript was written by T.L. and N.P.O. with numerous inputs from T.H.H. and L.F. The high-quality crystal was grown by J.J.I. and S.N. The basic characterization of crystals was made by J.J.I. and S.N. All authors discussed the results and commented on the manuscript.

\vspace{5mm}\noindent
{\bf Author Information} The authors declare no competing financial interests. Correspondence and requests for data and materials should be addressed to T.L. (liang16@stanford.edu) or N.P.O. (npo@princeton.edu).


\newpage

\clearpage
\newpage

\renewcommand{\thefigure}{S\arabic{figure}}
\renewcommand{\thesection}{S\arabic{section}}
\renewcommand{\theequation}{S\arabic{equation}}

\setcounter{equation}{0}
\setcounter{figure}{0}
\setcounter{table}{0}
\setcounter{section}{0}

\vspace{4mm}
{\bf Supplementary Information\\
} 
\vspace{6mm}

	\section{Group Theoretical Considerations on Octupolar Order \label{Tim}}
	
	As discuessed in sections A, B of method section, the observed orthogonal magnetization $M_\perp = \omega_{ijk} H_j H_k$ is the thermodynamical manifestation of the octupolar order $\omega_{ijk}$. 
	In this section, we use symmetry constrains and group theoretical considerations to derive the specific form of the octupolar order $\omega_{ijk}$. We pay attention to the experimental fact that the magnetically ordered state of Eu$_2$Ir$_2$O$_7$ has ordering vector $\bm{q} = 0$~\cite{Nakatsuji2013}.  Depending on the spin (local dipole) arrangements on a tetrahedron unit cell, pyrochlore magnets can display a variety of ${\bf q}=0$ magnetic orders. While the sum of local dipoles $\bm{m}_i$ is zero, and the dipolar order $\bm{M}^d = \sum_1^N\bm{m}_i = 0$ vanishes, the configuration of local dipoles gives the symmetry constraints on the octupolar order $\omega_{ijk}$ that is allowed. Unlike the dipolar order $\bm{M}^d$ which is a vector, multipolar orders are mathematically described by high-rank tensors that specify their transformation laws under point group symmetry operations acting jointly on the lattice and spin.  On each tetrahedron, the $12-3=9$ degrees of freedom for ${\bf q}=0$ magnetic order decompose into four different representations of the pyrochlore point group: $A_2, E, T_1,T_2$.  We will focus on $A_2$, a one-dimensional representation corresponding to all-in-all-out magnetic order (shown in Fig. 1a), and $T_2$, a three-dimensional representation corresponding to the order shown in Fig. 1b. 
	
	The all-in-all-out magnetic order is identified with a rank-3 tensor $\omega_{ijk}$, with $\omega_{abc} = \omega_{acb}= \omega_{bac} = \omega_{bca} = \omega_{cab} = \omega_{cba}\equiv \omega$ and all other components being zero, where $a,b,c$ are the three cubic axes. $\omega_{ijk}$ transforms identically as the all-in-all-out order: for example, it is invariant under three-fold rotation along the $(111)$ axis and changes sign under the two-fold rotation which sends $b\rightarrow-b, c\rightarrow -c$. 
	
	The nonlinear magnetization of this state can be deduced by considering the free energy under an applied field $\bf H$. A unique term involving the product of the $A_2$ order parameter $\omega$ and 
	third order polynomials of $\bf H$ is symmetry allowed: 
	\be
	F = \chi_\perp \omega H_a H_b H_c
	\ee
	
	Taking the derivative with respect to $H_c$ yields the orthogonal magnetization:
	\be
	M_{\perp}= \chi_\perp \omega H_a H_b 
	\label{A2}
	\ee
	This gives rise to the $d$-wave signal observed when field-cooled along $y$-axis.
	
	On the other hand, the $T_2$ order parameter has three components denoted by the multiplet $(t_a, t_b, t_c)$.  Using the fact that third-order polynomials of $H$ with the $T_2$ symmetry have the form $(H_a (H_b^2 - H_c^2), H_b (H_c^2 - H_a^2), H_c (H_a^2 - H_b^2))$, we can form a scalar in the free energy with the following coupling:
	\begin{eqnarray}
	F = && \tilde{\chi}_\perp \left[ t_a H_a (H_b^2 - H_c^2)  \right. \nonumber\\
	&  &  				\left.  + t_b H_b (H_c^2 - H_a^2)+ t_c H_c (H_a^2 - H_b^2) \right].  
	\end{eqnarray} 
	
	The derivative of $F$ with respect to $\bf H$ yields the magnetization to second order in the applied field: 
	\be
	M_c =\tilde{\chi}_\perp  \left[ t_c (H_a^2 - H_b^2)  + 2 t_b H_b H_c  - 2 t_a H_a H_c \right]. \label{t2-m}
	\ee
	$M_a$ and $M_b$ are obtained from $M_c$ by permutation of indices.  
	
	$M_c$, depicted in the last basis configuration of Fig. 1, gives rise to the shifted $d$-wave signal observed when field-cooled along $x$-axis.

	\begin{figure}
		\centering
		\includegraphics[width=9cm]{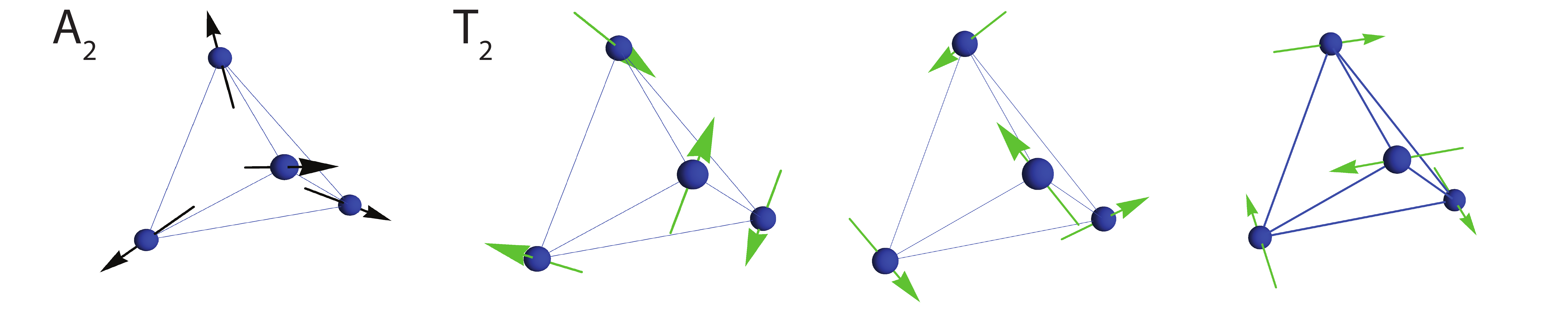} \label{A2,T2}
		\caption{(a) All-in-all-out order transforms in the $A_2$ representation (b) The three basis configurations of $T_2$ order.  The far right configuration gives rise to the shifted $d$-wave pattern observed when field-cooling along $x$ direction.}
	\end{figure}

	\section{Crystal growth}
	
	
	Single crystals of Eu$_2$Ir$_2$O$_7$ were grown by a KF flux method from a polycrystalline sample prepared by solid-state reaction of the appropriate mixture of Eu$_2$O$_3$ and IrO$_2$ powders (both of 4-nines purity). Using both powder and single-crystal X-ray diffraction measurements, we confirmed the growth of a single-phase crystal with the pyrochlore structure with lattice constant $a =$ 10.27 \AA. The single crystal has a natural growth plane normal to the [111] direction. The (111) planes meet along the [110] axis. For details, we refer the reader to Ref.~\cite{Nakatsuji2007}.
	
	
	\section{Experimental Details}
\begin{figure*}[h]
		\includegraphics[width=15 cm]{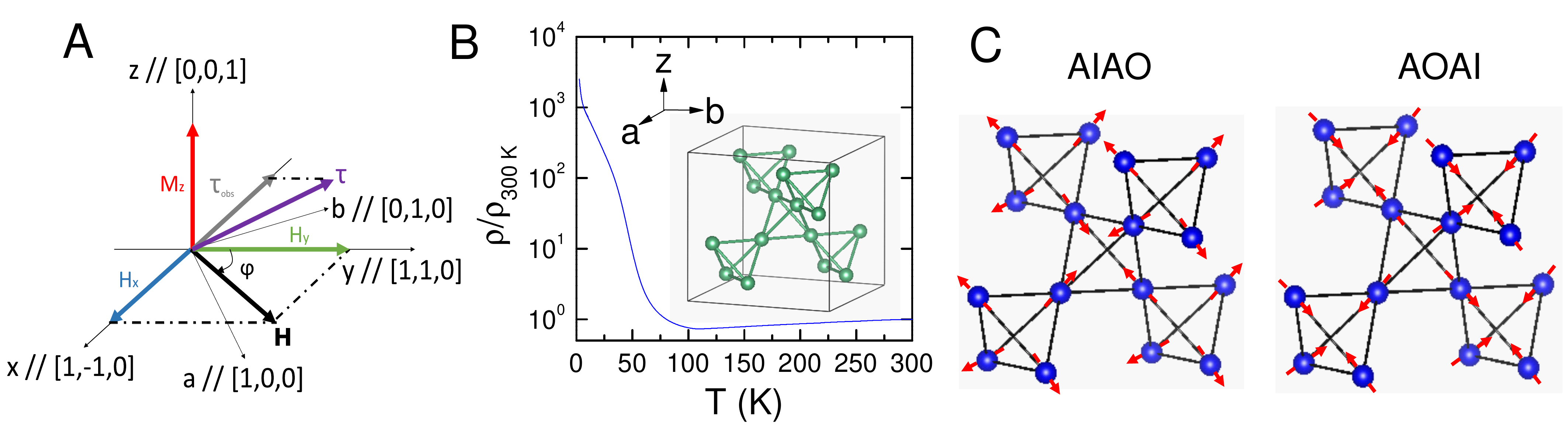}
		\caption{\label{Pyrochlore} 
			(A) Schematic of the orthogonal magnetization observed by torque magnetometry. The magnetic field {\bf H} lies in the $a$-$b$ plane at an angle $\varphi$ to the $y$-axis ($\equiv $[1,1,0]-axis). The cantilever detects only the torque component parallel to $x$-axis ($\equiv $[1,-1,0]-axis). The total observed torque magnetization $M_z$ is defined by $\tau_{obs} = M_zH\cos\varphi$.
			(B) The temperature dependence of the resistivity ratio $\rho/\rho_{300 K}$. $\rho$ shows a metal-to-insulator transition at $T_N\sim$ 120 K. The inset shows the Ir atoms in the unit cell of Eu$_2$Ir$_2$O$_7$. The Ir atoms define a network of corner-sharing tetrahedra, which are arranged in the diamond configuration within a pyrochlore unit cell.
			(C) Schematic of the AIAO (all-in-all-out) state and AOAI (all-out-all-in) state in the pyrochlore unit cell. For clarity, the boundary of the unit cell is not shown. In each tetrahedron, the spins at the vertices either all point ``in'' or``out''. AOAI and AIAO are time-reversed partners.  
		}
	\end{figure*}

	The pyrochlore structure of Eu$_2$Ir$_2$O$_7$ is comprised of a network of corner-sharing tetrahedra, as shown in the inset of Fig. \ref{Pyrochlore}B. The profile of the resistivity $\rho$ vs. temperature $T$ shows a metal-to-insulator transition at T$_N\sim $ 120 K, below which the system orders magnetically (Fig. \ref{Pyrochlore}B). A candidate for the magnetically ordered state is the all-in-all-out state (AIAO) in which the 4 spins at the vertices of a tetrahedron either point in or point out, as shown in Fig \ref{Pyrochlore}C. The time-reversed partner of the AIAO state is the AOAI state.
	
	In our torque set-up, the cantilever is made of a 10 $\mu$m-thick gold foil of length $\sim $ 5 mm and width $\sim $ 0.65 mm. The sample is glued to the cantilever with the [1,-1,0]-axis and the [1,1,-2]-axis aligned parallel to the short and long directions, respectively. The [1,1,1]-axis is perpendicular to the cantilever plane (Fig. \ref{Pyrochlore}A). 
	
	The cantilever only detects the $x$-component of the torque vector. We define the observed signal as ${\bf \tau}_{obs} \equiv -\vec{\tau}{\bf\cdot \hat{x}}$. The cantilever plane is aligned with the $z$ axis using a jig with wedge angle $\theta = \arctan(1/\sqrt{2})\sim 35^\circ$, as shown in Fig. \ref{fighyst}E. To vary the angle $\varphi$ between $\bf H$ and $\hat{\bf y}$, the sample is rotated about the $z$-axis (see Fig. \ref{fighyst}D,E). To investigate the hysteretic behaviors, we apply the field ${\bf H}_{fc}$ (fixed at 9 T) at 300 K, and cool the system to 5 K before performing measurements.

	\section{Resolving $M_\perp$ from $M_s$ and $M_p$ \label{ResM}}
	The observed torque is written as 
	\begin{eqnarray}
	\tau_{obs}  &=& M_zH\cos\varphi   \nonumber\\
	& = & (M_s+M_p+M_{\perp}) H\cos\varphi \nonumber\\
	& = & \tau_s+\tau_p+\tau_{\perp},
	\label{tau}
	\end{eqnarray}
	where $\tau_s=M_sH\cos\varphi$ $(\propto H)$, $\tau_p=M_pH\cos\varphi$ $(\propto H^2)$, $\tau_{\perp}=M_{\perp}H\cos\varphi$ $(\propto H^3)$. The $H$-odd and $H$-even components are $\tau_{odd} = \tau_s + \tau_\perp$ and $\tau_{even} = \tau_p$.
	
\begin{figure*}[h]
		\includegraphics[width=15 cm]{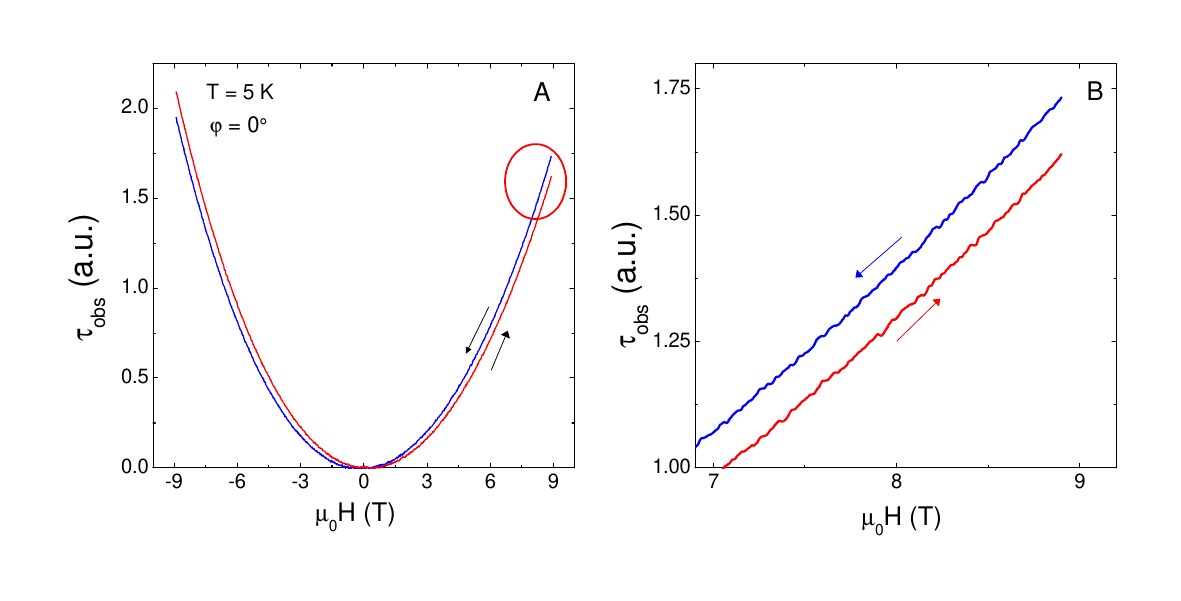}
		\includegraphics[width=15 cm]{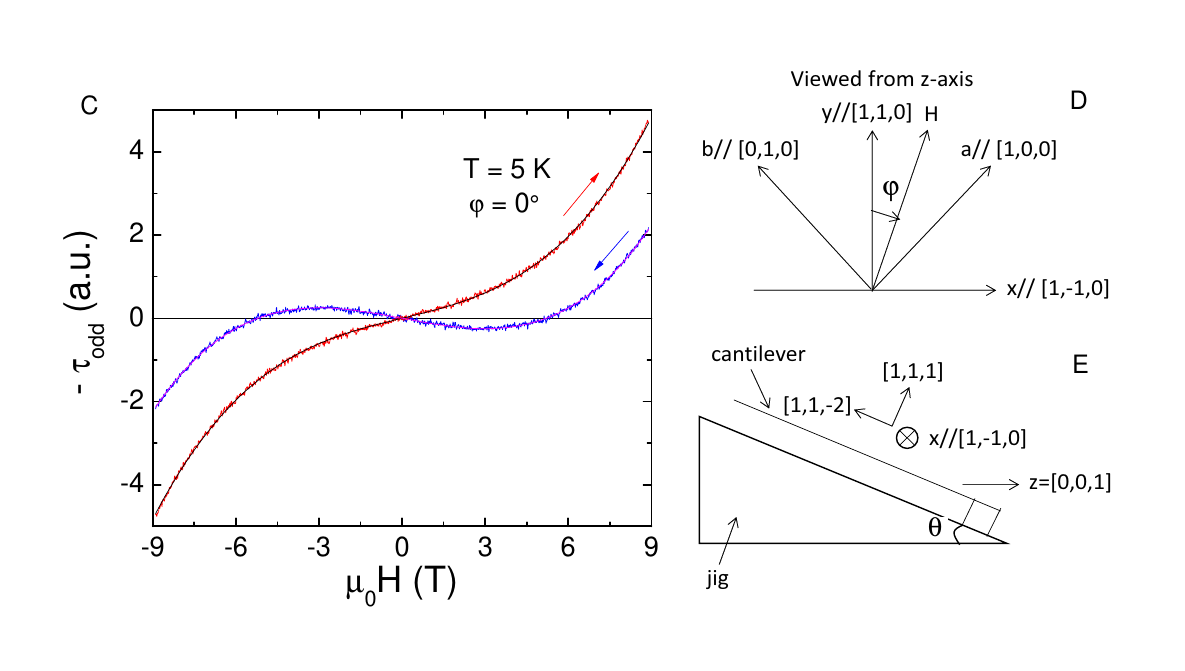}
		\caption{\label{fighyst} 
			Hysteretic behavior of the observed torque measured at $T$ = 5 K and tilt angle $\varphi = 0$.
			Panel (A) plots the total observed torque $\tau_{obs}$ vs. $H$ for sweep-up (red) and sweep-down (blue) field scans. The high-field region circled in red is shown magnified in Panel (B). An abrupt jump occurs when the sweep direction is reversed at $H$ = 9 T. Panel (C) shows the $H$-odd component $\tau_{odd}$ of the trace in Panel (A) after antisymmetrization. $\tau_{odd}$ is the sum $\tau_\perp + \tau_s$. The hysteresis is identified as arising entirely from $\tau_s\sim H$, whereas $\tau_\perp\sim H^3$ is reversible. Fits to the polynomial $c_1 H + c_3 H^3$ are shown as the thin black and magenta curves. Panels (D) and (E) are schematics of the set-up for torque magnetometry. The crystal was mounted on the cantilever with its [1,-1,0]-axis ([1,1,-2]-axis) parallel to the short (long) directions of the cantilever. The [1,1,1]-axis is perpendicular to the cantilever. A jig with wedge angle $\theta = \arctan(1/\sqrt{2})\sim 35^\circ$ was employed to keep the cantilever plane fixed at the angle $\theta$ to the $z$ axis. The cantilever detects only the component of $\vec\tau$ along the $x$ axis.
		}
	\end{figure*}

	For illustration, we show in Fig. \ref{fighyst}A the observed $\tau_{obs}$ measured at 5 K with $\varphi$ fixed at 0$^\circ$. The dominant contribution comes from the background paramagnetic term $\tau_p\sim H^2$. The asymmetry of the curves reveals the presence of a finite $\tau_{odd}$ that is hysteretic (shown in expanded scale in Fig. \ref{fighyst}B). By antisymmetrizing the curves in Panel A, we obtain the curves of $\tau_{odd}$ plotted in Fig. \ref{fighyst}C. As shown, both branches of $\tau_{odd}$ vs. $H$ can be closely fitted to the polynomial $c_1H + c_3H^3$ (fits shown as thin black and magenta curves). The two terms are identified with $\tau_s$ and $\tau_\perp$, respectively. At 5 K, $\tau_\perp$ is nearly identical for both branches, i.e. all the hysteretic behavior resides in $\tau_s$. The temperature and angular dependence of hysteretic behavior between $\tau_s$ and $\tau_\perp$ are totally different as shown in Figs.~\ref{[1,1,0]-Full},~\ref{[1,-1,0]-Full} (see Sec. \ref{angular}). 
	
	\section{Angular dependence and temperature dependence}\label{angular}
	
	We provide a detailed discussion of the angular dependence of the three torque components and their hysteretic behavior vs. $T$. We illustrate the important influence of the field ${\bf H}_{fc}$ in which the sample is cooled from above $T_N$ to 5 K.
	
	\subsection{Angular dependence and hysteresis}
	Figure \ref{[1,1,0]-Full} shows two groups of panels. The left panels (group A) 
	refer to measurements performed after field cooling in the field ${\bf H}_{fc} \parallel -{\bf \hat{y}}$, while the right panels (group B) show the same measurements performed after cooling with ${\bf H}_{fc}$ in the reversed direction (+${\bf \hat{y}}$).
	
	We will focus on the group A panels. The Panel Aa shows the angular variation of the torque component $\tau_{\perp}$ associated with the orthogonal magnetization $M_\perp$. The thin red curve is a fit to the expression $\tau_\perp = M_{\perp} H\cos\varphi \propto H^3$ with $M_\perp$ given by Eq.~9 in the main text. Red (blue) circles are for field sweep-up (sweep-down) curves while black circles denote their average. The component $\tau_\perp$ at 5 K is reversible when $H$ is swept between $\pm$9 T for all values of $\varphi$ (red and blue circles coincide within our measurement uncertainty). The inset shows the $d$-wave pattern with positive lobe along the $+y$ axis. 
	
	Panel \ref{[1,1,0]-Full}Ab shows the $T$ dependence of $M_\perp$ measured at fixed $\varphi = 180^\circ$ in sweep-up (red circles) and sweep-down (blue) curves as $T$ is varied between 5 and 150 K (black circles are their average). As $T$ is raised above 25 K, the sweep-up points (red circles) deviate strongly from the sweep-down points (blue). The hysteresis amplitude $\Delta M_\perp$ (defined as the difference between the two curves) rises rapidly to a maximum near 70 K in a thermally activated way (this is discussed in more detail below, in Fig. \ref{fighystamp}). Between 90 and 120 K, the amplitude reverses in sign. Above $T_N$, $M_\perp$ vanishes altogether within our resolution.
	
	Panel \ref{[1,1,0]-Full}Ac shows the angular dependence of the background paramagnetic term $\tau_p$ measured at 5 K. No hysteresis is resolvable. The fit to $\tau_p$ = $M_p H\cos\varphi$ with $M_p$ given by Eq.~11 (main text) is shown as the red curve. The angular pattern is dipolar as shown in the sketch. The data for $M_p$ measured at $\varphi = 180^\circ$ for sweep-up and sweep-down curves are plotted in Panel \ref{[1,1,0]-Full}Ad. The hysteresis amplitude $\Delta M_p$ is very small throughout. (We believe the small finite values within the interval 60$\to$100 K arise from errors in subtracting the contribution from $\Delta M_\perp$. See Figs. \ref{[1,-1,0]-Full}Ad and Bd below). 
	
	Panel \ref{[1,1,0]-Full}Ae shows the angular dependence of $\tau_s$ at 5 K. Unlike $\tau_\perp$, $\tau_s$ shows a very large hysteresis (red and blue circles indicate sweep-up and -down, respectively). We are not able to identify the unusual angular dependences with an analytic form. The $T$ dependence of $M_s$ measured at $\varphi = 180^\circ$ shows that the hysteresis amplitude is large below 60 K but becomes negligible as $T\to T_N$ (Panel Af). 
	
	The panels in group B of Fig. \ref{[1,1,0]-Full} are in exactly the same sequence as the panels in group A. When ${\bf H}_{fc}$ is reversed in sign, it causes both $M_\perp$ and $M_s$ to reverse in sign (the magnitudes remain unchanged from group A to our resolution). The signs of the lobes of the $d$-wave pattern (inset in Ba) are now reversed. The chirality (see Eq.~6 of main text) has the value $\omega = 1$ (right-handed) in Panel Aa (${\bf H}_{fc} \parallel-\bf\hat{y}$). In Panel Ba, however, $\omega$ equals -1 (left-handed) for ${\bf H}_{fc}\parallel \bf\hat{y}$. The chirality $\omega$ changes from +1 to -1 when ${\bf H}_{fc}$ is reversed.
	
	In contrast, the sign of the paramagnetic term $M_p$ is unaffected by the field reversal (the sign of the dipole pattern in Panel \ref{[1,1,0]-Full}Ac is the same as in Bc).

	Figure \ref{[1,-1,0]-Full} shows the same sequence of panels as Fig. \ref{[1,1,0]-Full}, but for the situation when the sample is cooled with the field ${\bf H}_{fc}\parallel \pm{\bf \hat{x}}$. The discussion closely follows that above. 
	The major difference is that the lobe direction of the $d$-wave pattern is now rotated by 45$^\circ$ relative to the previous pattern (see insets in Panels \ref{[1,-1,0]-Full}Aa and \ref{[1,-1,0]-Full}Ba). The $d$-wave pattern is now desribed by $M_{\perp}$ = $\chi^x_\perp H^2(1+\eta\cos\varphi)\sin 2\varphi$.
	
	For the orthogonal term, the hysteresis amplitude $\Delta M_\perp$ is large between 35 and 90 K, but again becomes unresolvably small below 25 K (Panels \ref{[1,-1,0]-Full}Ab and Bb). For the paramagnetic term, the hysteresis amplitude $\Delta M_p$ is unresolved from zero throughout the entire interval 5 $\to$ 150 K (Panels \ref{[1,-1,0]-Full}Ad and Bd). 
	
	Finally, when ${\bf H}_{fc}$ is reversed (compare groups A and B in Fig. \ref{[1,-1,0]-Full}), both $M_\perp$ and $M_s$ reverse their signs, but $M_p$ is unaffected.
	
	\subsection{Hysteresis amplitude $\Delta M_\perp$}
	Figure \ref{fighystamp} shows the $T$ dependence of the hysteresis amplitude of the orthogonal magnetization $\Delta M_\perp$ (difference between $M_\perp$ measured in the sweep-up and sweep-down branches). Panels \ref{fighystamp}A and B plot, in semilog scale, the values of $\Delta M_\perp$ measured after cooling in the field ${\bf H}_{fc} \parallel -{\bf \hat{y}}$ and $-{\bf \hat{x}}$, respectively. The data below 70 K fit well to a thermally activated form with gaps $\Delta$ = 170 K and 220 K, respectively. Panels \ref{fighystamp}C and D show in linear scale the $T$ dependence of $\Delta M_\perp$ up to 150 K. 
	
	A finite amplitude $\Delta M_\perp$ reflects the diffusion of domain walls separating domains with $d$-wave patterns of opposite signs in the non-equilibrium state created by a field reversal. The rapid decrease of $\Delta M_\perp$ below 70 K suggests that the diffusion rate is exponentially suppressed as $T\to$ 5 K. The reversible nature of the curves of $M_\perp$ (see Fig. 1D of main text) results from very strong pinning of the domain walls at low $T$. 
	
	By contrast, the domain walls separating domains of opposite sign of the field independent magnetization $M_s$ diffuse freely at 5 K, as evidenced by the large low-$T$ hysteresis amplitude shown in Figs. \ref{[1,1,0]-Full}Af and Bf.\\

	\subsection{[1,0,0]-axis, [0,1,0]-axis field cooling}
	Figure \ref{100_010} shows the angular dependence of the cubic term of torque $\tau_{\perp}$ = $M_{\perp} H\cos\varphi$ for field cooling along [1,0,0]-axis and [0,1,0]-axis. The curves can be fitted using the linear combination of that of [1,1,0]-axis ($y$-axis) field cooling and [1,-1,0]-axis ($x$-axis) field cooling, namely, $M_{\perp}$ = $(1+\eta\cos\varphi)(\chi_\perp^y\cos2\varphi+\chi_\perp^x\sin2\varphi)$ for [1,0,0]-axis field cooling and $M_{\perp}$ = $(1+\eta\cos\varphi)(\chi_\perp^y\cos2\varphi-\chi_\perp^x\sin2\varphi)$ for [0,1,0]-axis field cooling, showing [1,1,0]-axis and [1,-1,0]-axis are the principal axes.
	
	\subsection{[1,1,0]-axis, [1,-1,0]-axis field cooling, high temperatures}
	Figure \ref{110_1-10_65K} shows the angular dependence of the cubic term of torque $\tau_{\perp}$ = $M_{\perp} H\cos\varphi$ at 65K for field cooling along [1,1,0]-axis and [1,-1,0]-axis. Although the angular dependence of red and blue dots (sweeping field up and down) show complicated behavior due to the domain formation, the averaged curves fit to $M_{\perp}$=$\chi_\perp^A(1+\eta\cos\varphi)\cos2\varphi$ (for panel A) and $M_{\perp}$=$\chi_\perp^B(1+\eta\cos\varphi)\sin2\varphi$ (for panel B), suggesting partial ordering of the system.
	
	\subsection{Zero field cooling}
	Figure \ref{Zero_Field_Cooling} shows the angular and temperature dependence of the cubic term of torque $\tau_{\perp}$ = $M_{\perp} H\cos\varphi$ measured after zero-field cooling of the sample. We observe two sets of behavior. In one set (Set 1), the orthogonal magnetization $M_\perp$ has an angular variation similar to that obtained by field cooling with ${\bf H}_{fc}\parallel \pm {\bf\hat{y}}$ (but not that obtained with ${\bf H}_{fc}\parallel \pm {\bf\hat{x}}$). This suggests either a residual memory from an early field-cooled run, or that, when zero-field cooled, the domains spontaneously select the $d$-wave pattern in Fig. \ref{[1,1,0]-Full}Ab. In the second set of zero-field cooled runs (Set 2), the observed $M_\perp$ is close to zero. This suggests the formation of domains with opposite signs of nearly equal weights, so that the torque component $\tau_\perp$ mutually cancel nearly perfectly.

	\newpage
	
	\vspace{1cm}\noindent
	
	\vspace{1mm}
	$^\ast${These authors contributed equally to this work.}
	
	\vspace{3mm}
	$^\dagger${Current address of T.L.: Department of Applied Physics, Stanford University, Stanford, CA, 94305}
	
	$^{\dagger\dagger}${Current address of T.H.H.: Kavli Institute for Theoretical Physics, University of California, Santa Barbara, CA 93106}
	
	\vspace{3mm}\noindent
	{\bf Author Information} The authors declare no competing financial interests. Correspondence and requests for data and materials should be addressed to T.L. (liang16@stanford.edu) or N.P.O. (npo@princeton.edu).
	
	\begin{figure*}[h]
		\includegraphics[width=8.5cm]{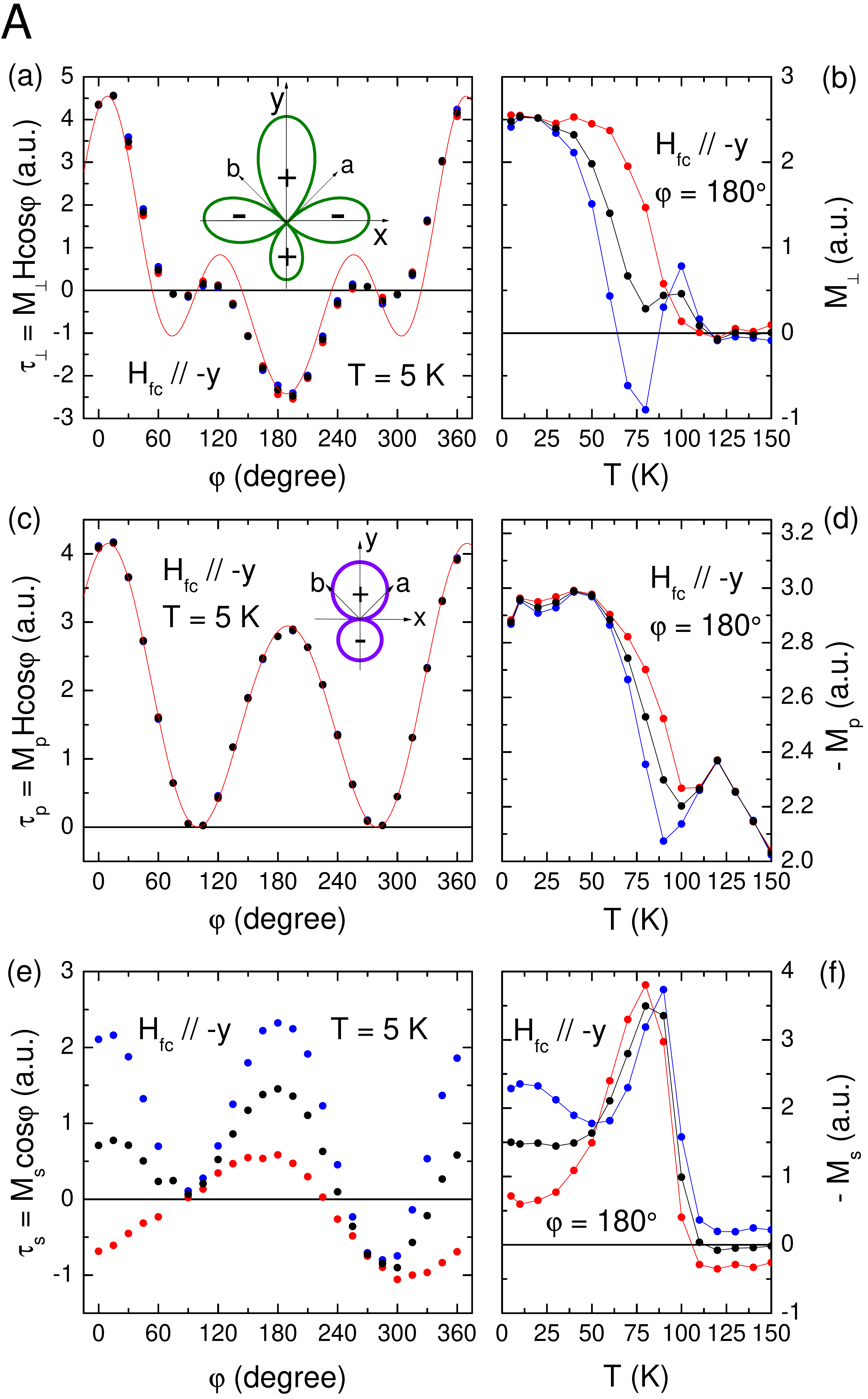}
		\includegraphics[width=8.5 cm]{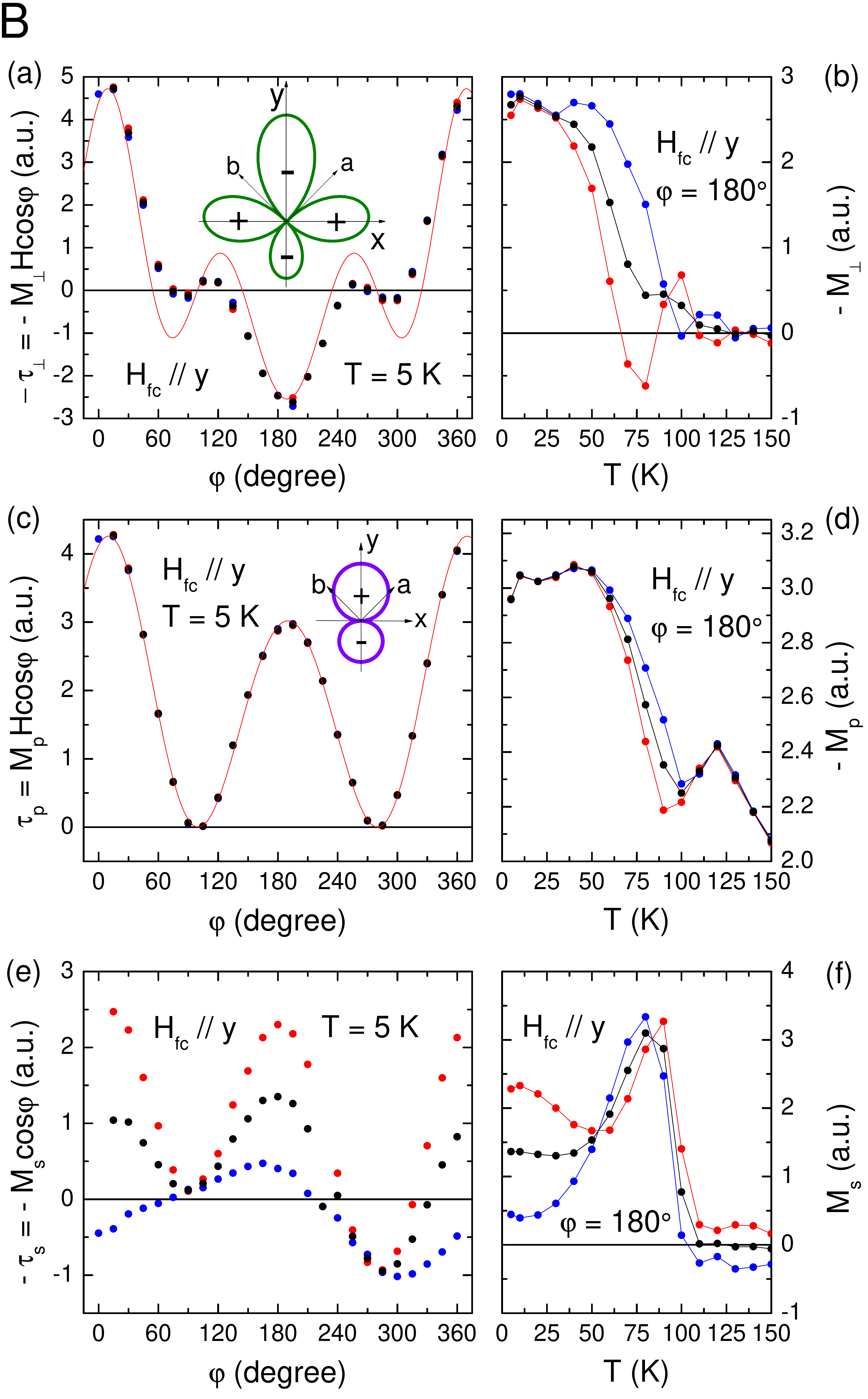}
		\caption{\label{[1,1,0]-Full} 
			The angular dependence of each of the 3 torque components $\tau_\perp$, $\tau_p$ and $\tau_s$ and their hysteretic amplitude versus $T$ observed when field-cooled in the field ${\bf H}_{fc}\parallel -{\bf\hat{y}}$ (Panels in A) and $\parallel {\bf\hat{y}}$ (Panels in B). Panel Aa shows the $d$-wave like variation of $\tau_\perp$ vs. $\varphi$ at 5 K. The thin red curve is a fit to Eq. 9 of main text. The $d$-wave pattern varies with $\varphi$ as $\cos 2\varphi$ with positive lobe on the $y$ axis (inset). Panel (Ab) shows the hysteresis vs. $T$ with $\varphi$ fixed at 180$^\circ$. The hysteresis amplitude is largest near 70 K, but becomes unobservable below 25 K. Panel (Ac) shows the trace of the paramagnetic term $\tau_p$ vs. $\varphi$ at 5 K. The fit (red curve) to Eq. 11 (main text) has the skewed dipolar form sketched in the inset. Panel (Ad) shows that the hysteresis is nearly zero except for $T$ between 60 and 100 K, where the small contribution comes from $\tau_\perp$ due to imperfect subtraction. Panel (Ae) shows $\tau_s$ vs. $\varphi$ at 5 K showing a pronounced difference between sweep-up (red circles) and sweep-down (blue) measurements. Panel (Af) plots the magnitude of $\tau_s$ vs. $T$, which shows a large hysteresis below 60 K, a pattern opposite to that of $\tau_\perp$. When field cooled with ${\bf H}_{fc}$ reversed ($\parallel {\bf \hat{y}}$), the curves of $\tau_p$ are nominally unchanged (Panels Bc and Bd). However, the curves for $\tau_\perp$ (Panels Ba and Bb) and $\tau_s$ (Be and Bf) reverse their signs.
		}
	\end{figure*}

	\begin{figure*}[h]
		\includegraphics[width=8.5 cm]{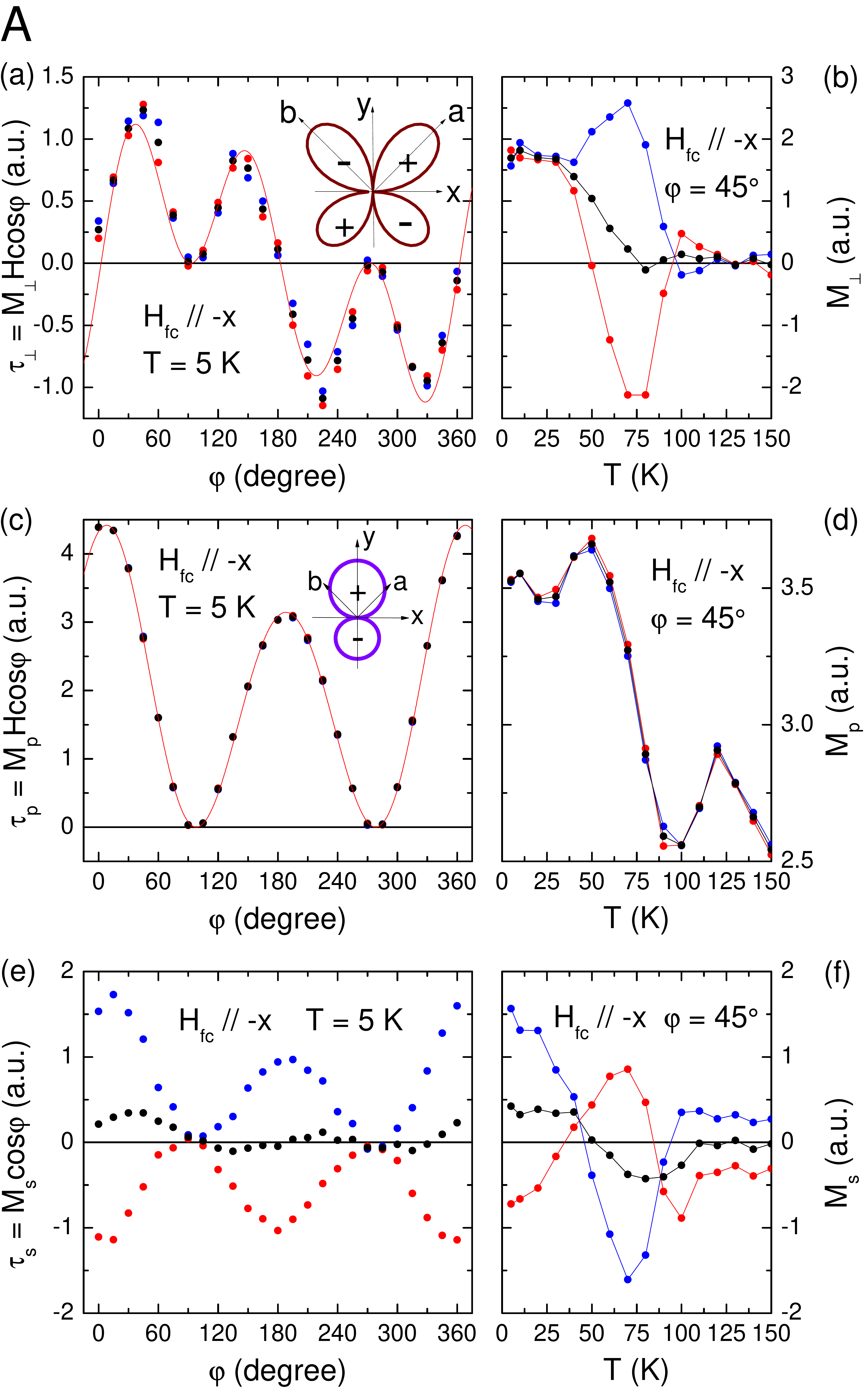}
		\includegraphics[width=8.5 cm]{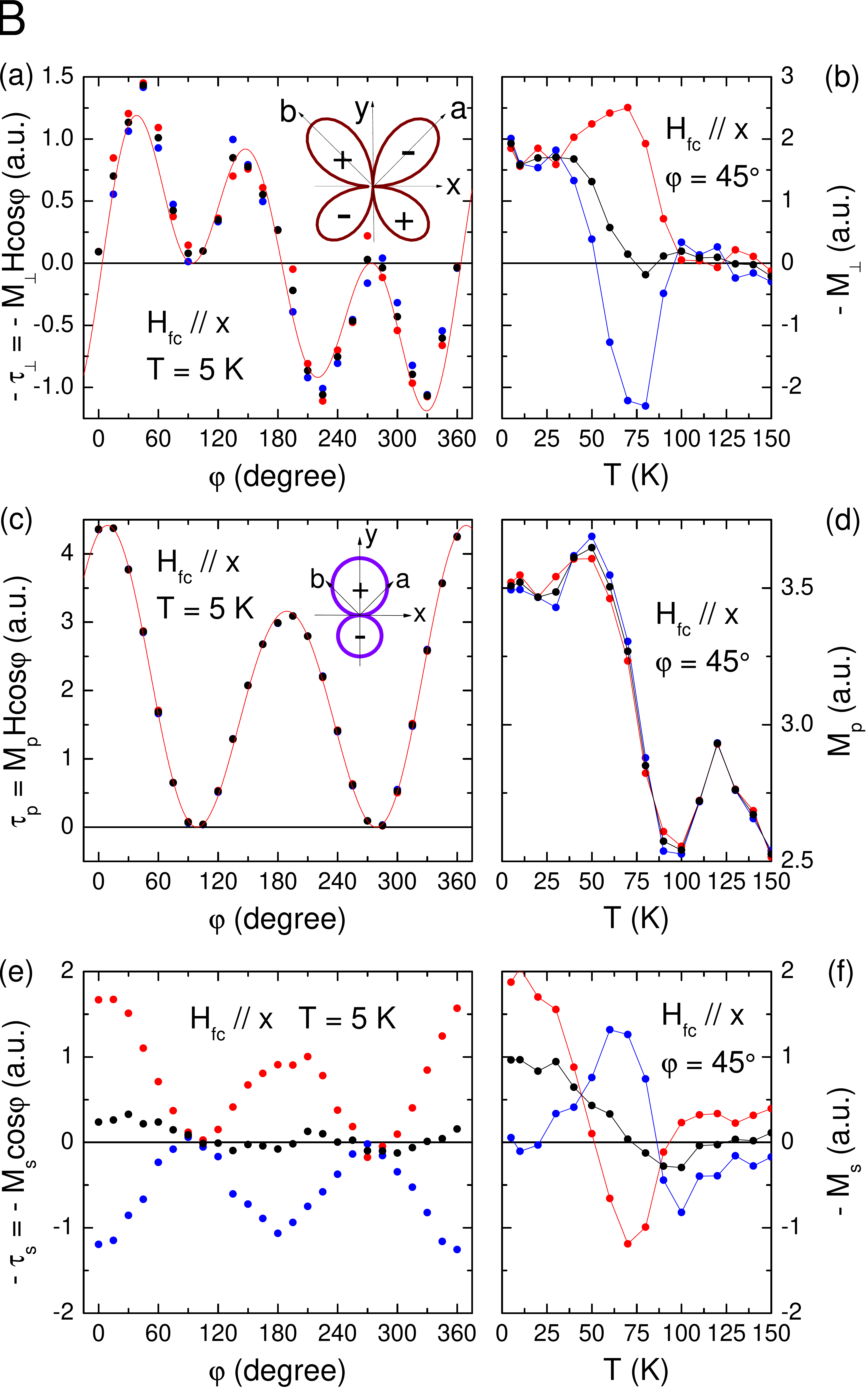}
		\caption{\label{[1,-1,0]-Full} 
			The angular dependence of the 3 torque components $\tau_\perp$, $\tau_p$ and $\tau_s$ and their hysteretic amplitude versus $T$ observed when field-cooled in the field ${\bf H}_{fc}\parallel -{\bf\hat{x}}$ (Panels in A) and $\parallel {\bf\hat{x}}$ (Panels in B). The description here is closely similar to that in Fig. \ref{[1,1,0]-Full} apart from the 90$^\circ$ rotation in ${\bf H}_{fc}$. Panel Aa shows the $d$-wave like variation of $\tau_\perp$ vs. $\varphi$ at 5 K. The thin red curve is a fit to Eq. 10 of main text. The $d$-wave pattern varies with $\varphi$ as $\sin 2\varphi$ with positive lobe on the $a$ axis (inset). Panel (Ab) shows the hysteresis vs. $T$ with $\varphi$ fixed at 45$^\circ$. The hysteresis amplitude is largest near 70 K, but becomes unobservable below 30 K. Panel (Ac) shows the trace of the paramagnetic term $\tau_p$ vs. $\varphi$ at 5 K. The fit (red curve) to Eq. 11 (main text) has the skewed dipolar form sketched in the inset. Panel (Ad) shows that the hysteresis amplitude is zero. Panel (Ae) shows $\tau_s$ vs. $\varphi$ at 5 K showing a pronounced difference between sweep-up (red circles) and sweep-down (blue) measurements. Panel (Af) plots the magnitude of $\tau_s$ vs. $T$, which shows a large hysteresis up to $T_N$. When field cooled with ${\bf H}_{fc}$ reversed ($\parallel {\bf \hat{x}}$), the curves of $\tau_p$ are nominally unchanged (Panels Bc and Bd). However, the curves for $\tau_\perp$ (Panels Ba and Bb) and $\tau_s$ (Be and Bf) reverse their signs.
		}
	\end{figure*}

	
	\begin{figure*}[h]
		\includegraphics[width=15 cm]{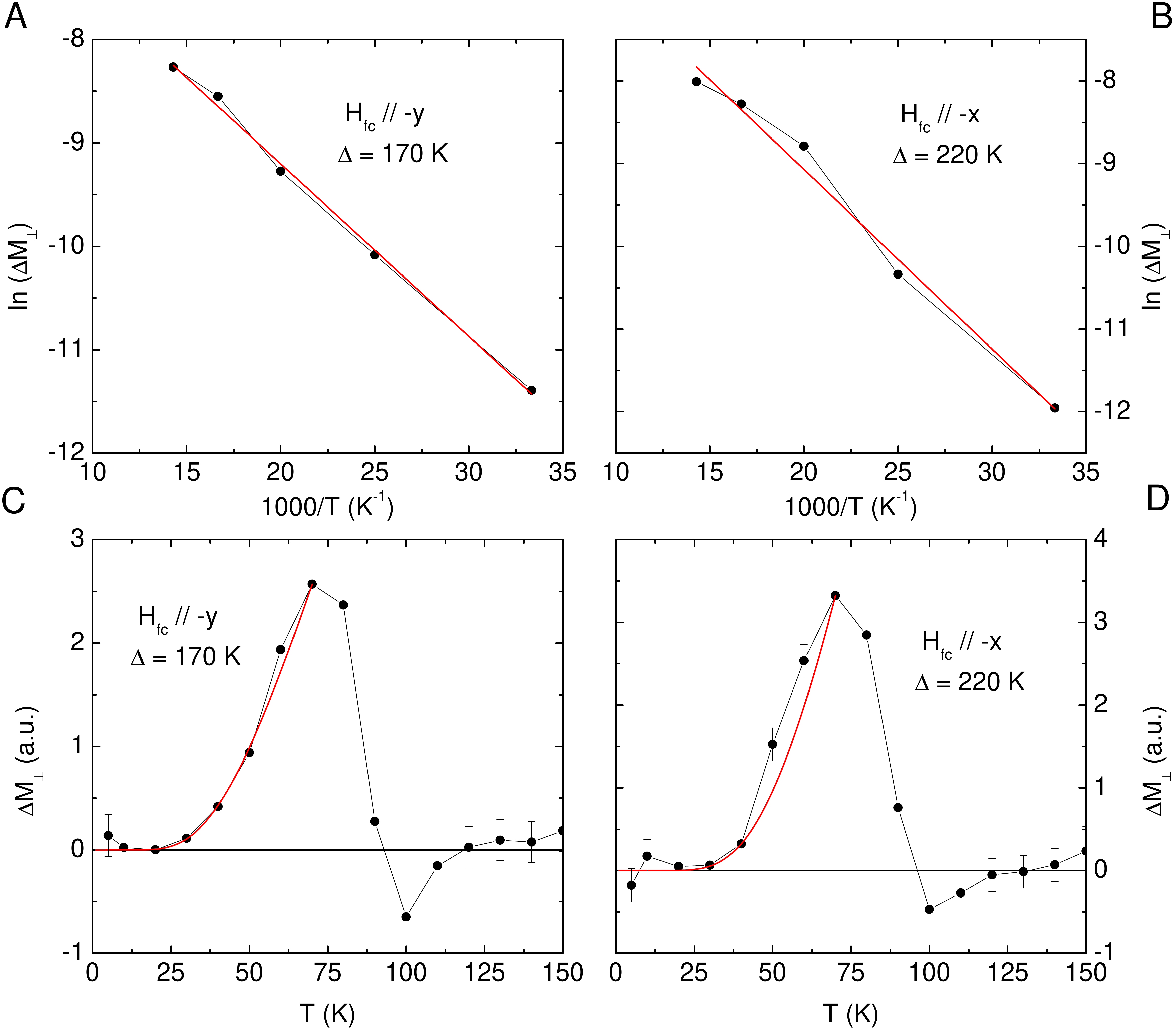}
		\caption{\label{fighystamp} 
			The $T$ dependence of the hysteresis amplitude $\Delta M_\perp$ (difference between $M_\perp$ measured in sweep-up and sweep-down curves). The semilog plot in Panel (A) shows that, below 70 K, $\Delta M_\perp$ is thermally activated with gap $\Delta\sim$ 170 K (observed with ${\bf H}_{fc}\parallel -{\bf \hat{y}}$). The gap is larger (220 K) when field cooled with ${\bf H}_{fc}\parallel -{\bf \hat{x}}$ (Panel B). Panels C and D show plots of $\Delta M_\perp$ vs. $T$ for the two directions of ${\bf H}_{fc}$ displayed in linear scale. The red curves are plots of the thermally activated form $\sim\exp{(-\Delta/T)}$. 
		}
	\end{figure*}
	

	
	\begin{figure*}[h]
		\includegraphics[width=15 cm]{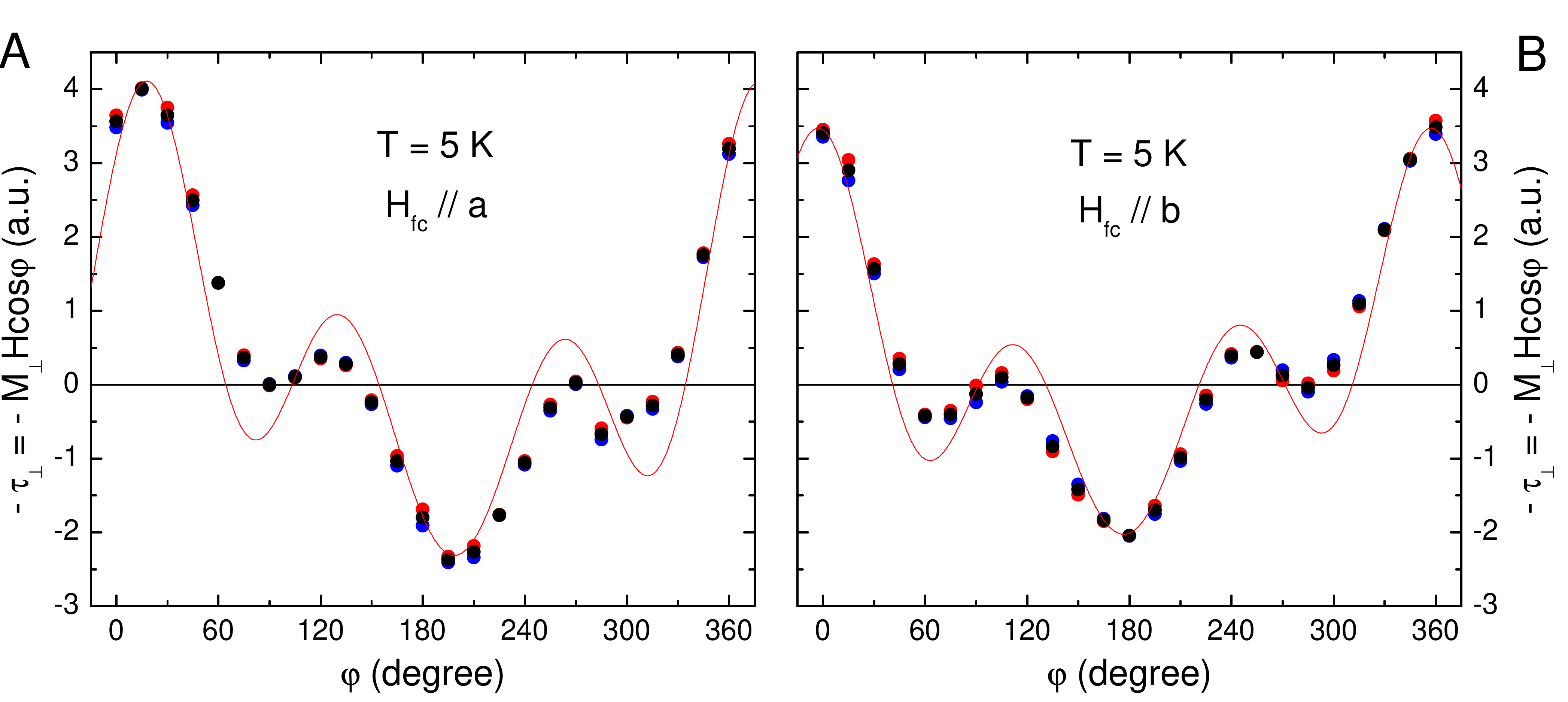}
		\caption{\label{100_010} 
			Angular dependence of cubic term of torque $\tau_{\perp}$ = $M_{\perp} H\cos\varphi$ under field cooling along [1,0,0]-axis (Panel A) and [0,1,0]-axis (Panel B). Red and blue solid circles represent data taken in field sweep-up and field-down scans, respectively. Black circles are their average.
		}
	\end{figure*}
	

	
	\begin{figure*}[h]
		\includegraphics[width=15 cm]{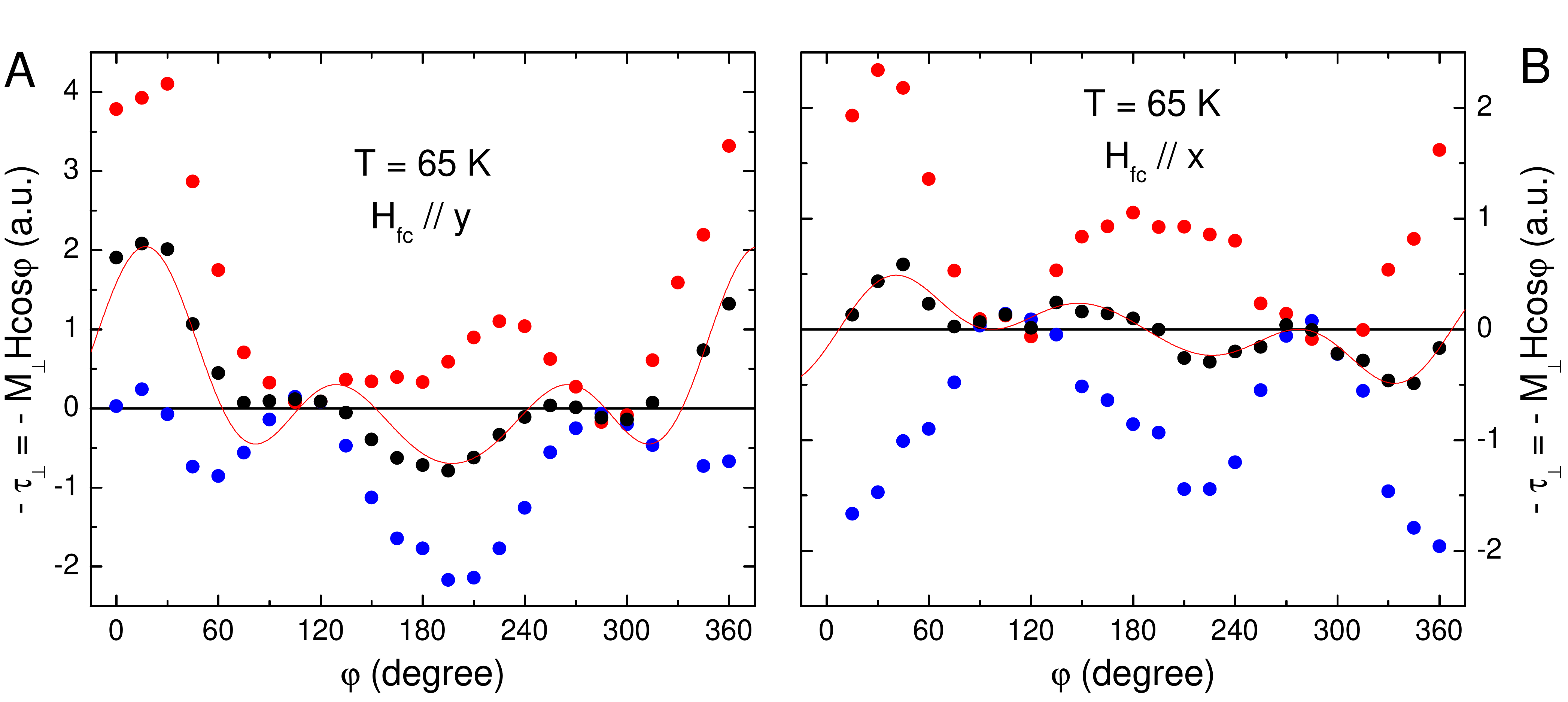}
		\caption{\label{110_1-10_65K}
			Angular dependence of cubic term of torque $\tau_{\perp}$ = $M_{\perp} H\cos\varphi$ under field-cooling with field along [1,1,0]-axis ($y$-axis) (panel A) and [1,-1,0]-axis ($x$-axis) (Panel B) at 65K. Red and blue solid circles represent data taken in field sweep-up and field-down scans, respectively. Black circles are their average.
		}
	\end{figure*}
	

	
	\begin{figure*}[h]
		\includegraphics[width=15 cm]{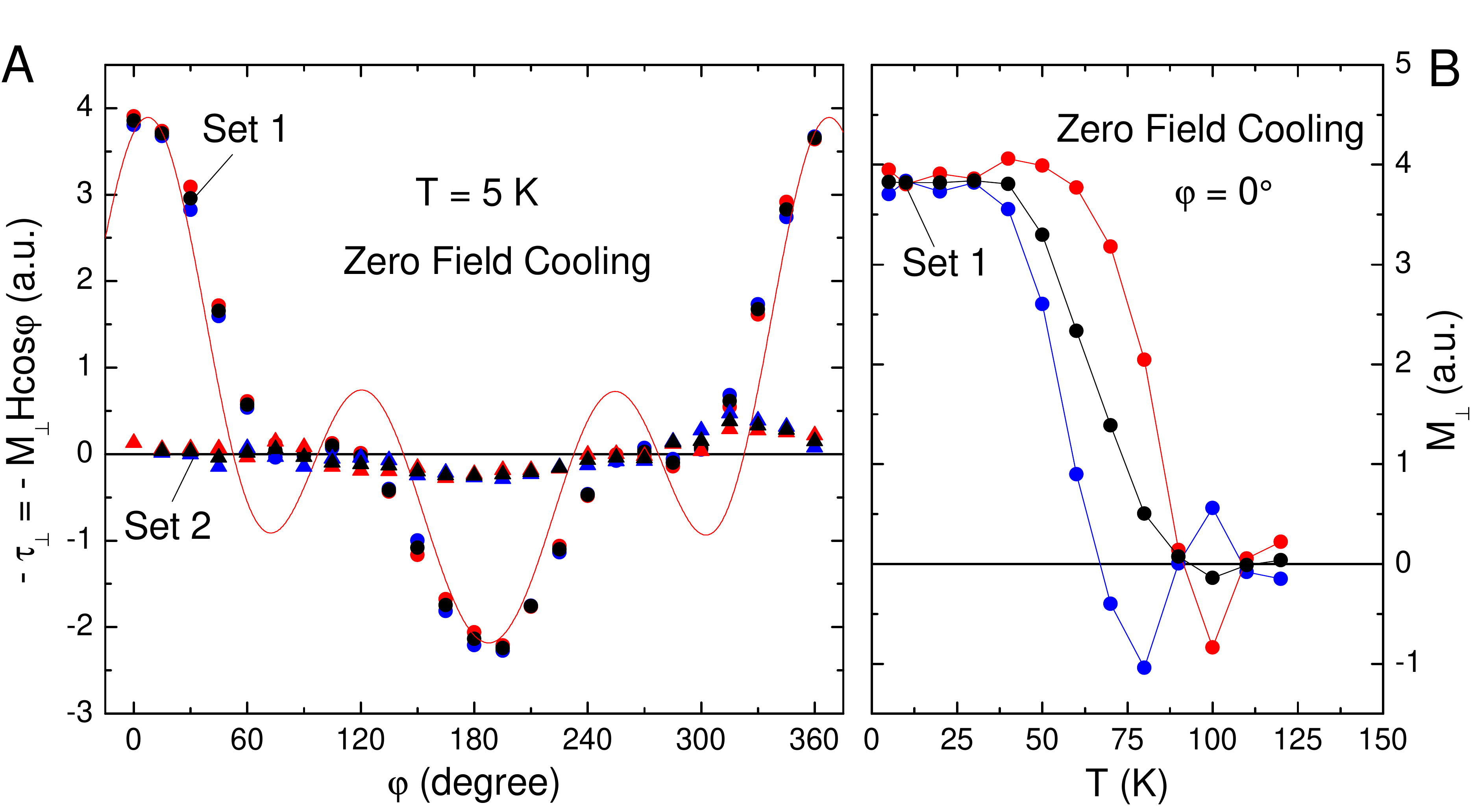}
		\caption{\label{Zero_Field_Cooling}
			Angular (panel A) and temperature dependence (panel B) of cubic term of torque $\tau_{\perp}$ = $ M_{\perp} H\cos\varphi$ under zero-field cooling. We observe two sets of behavior. In one (Set 1), the angular variation of $M_\perp$ is similar to that in Fig. \ref{[1,1,0]-Full}Ab. This suggests that the sample retains some memory of an earlier field-cooled run. In the other set (Set 2), the observed $\tau_\perp$ is very small, suggesting that the domain volumes with opposite signs are nearly equal in weight. This results in nearly complete cancellation of the orthogonal magnetization. Red and blue solid circles represent data taken in field sweep-up and sweep-down scans, respectively. Black circles are their average.
		}
	\end{figure*}
	
	
	
	\begin{figure*}[h]
		\includegraphics[width=15 cm]{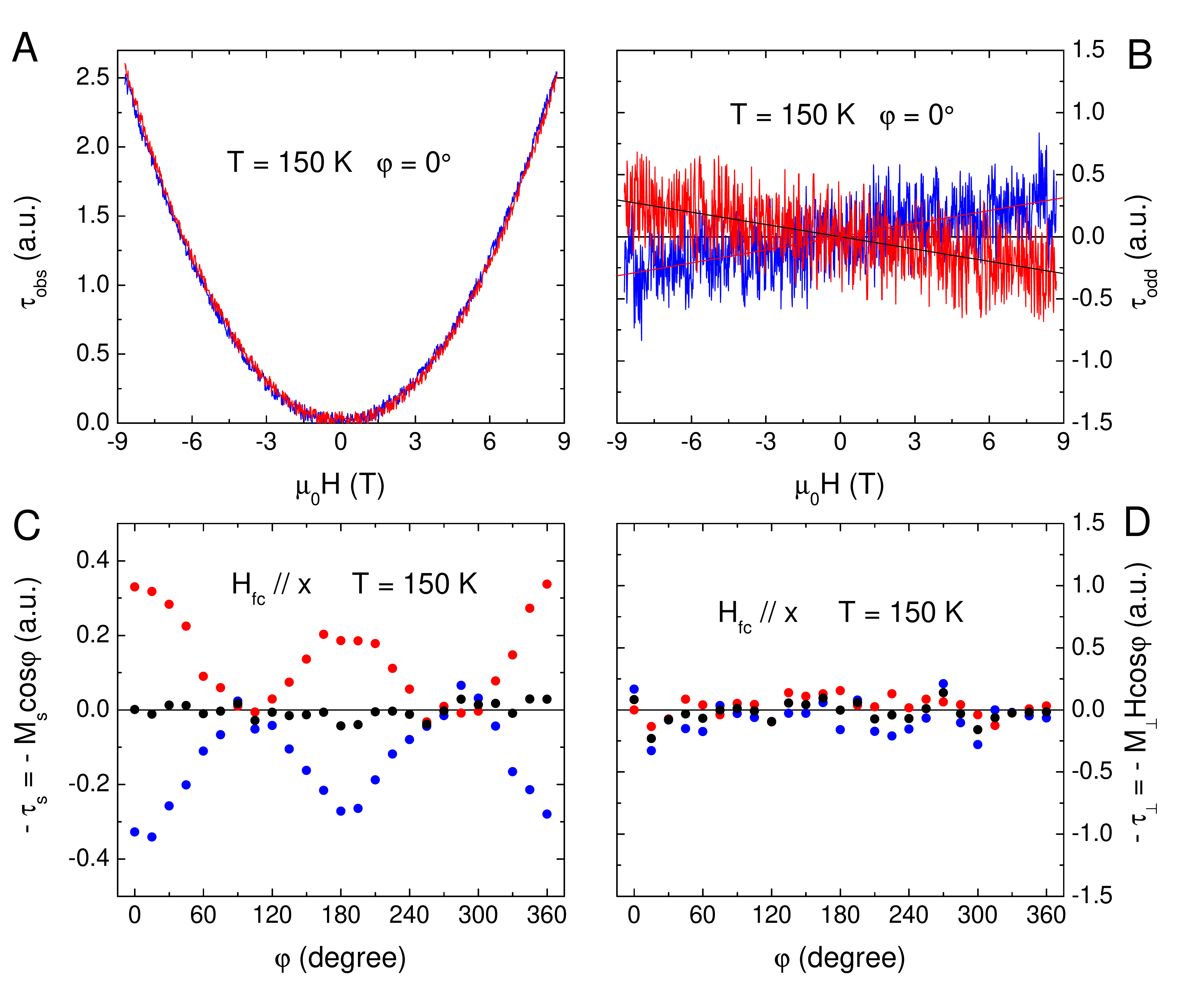}
		\caption{\label{150K} 
			Hysteretic behavior of the observed torque measured at $T$ = 150 K ($>T_N$ = 120 K) and tilt angle $\varphi = 0$.
			Panel (A) plots the total observed torque $\tau_{obs}$ vs. $H$ for sweep-up (red) and sweep-down (blue) field scans. Parabolic behavior coming from paramagnetic term M$_p$ dominates the signal with small tilt coming from contribution of M$_s$ term. Panel (B) shows the $H$-odd component $\tau_{odd}$ of the trace in Panel (A) after antisymmetrization. Fits to the polynomial $c_1 H + c_3 H^3$ are shown as the thin black and magenta curves. c$_3$ vanishes reflecting that no ${\bf M}_\perp$ term exists above $T_N$ = 120 K. Panel (C) plots the angular dependence of $\tau_s = M_s H \cos\varphi$. M$_s$ persists above $T_N$, having different origin from ${\bf M}_\perp$. Panel (D) plots the angular dependence of $\tau_\perp = M_\perp H\cos\varphi$, which shows vanishing values, showing no ${\bf M}_\perp$ exists above $T_N$.   
		}
	\end{figure*}
	
	
	%
	%
\end{document}